\documentclass[a4paper,11pt]{article}
\usepackage{aaskaiid}
\usepackage{orcidlink}
\usepackage{multirow}
\usepackage{ulem}

\title{Cosmology from H{\sc\,i} galaxy surveys with the SKA}
\ShortTitle{Cosmology from H{\sc\,i} galaxy surveys with the SKA}

\author[1]{Ainulnabilah Nasirudin \orcidlink{0000-0003-2213-4547}}
\ShortName{A. Nasirudin et al.} 
\author[1,2]{Philip Bull \orcidlink{0000-0001-5668-3101}}
\author[3,4,5]{Ziad Sakr \orcidlink{0000-0002-4823-3757}}
\author[6]{Jo\"{e}l Mayor \orcidlink{0009-0008-6499-8013}}
\author[7,8,9]{Benedict Bahr-Kalus \orcidlink{0000-0002-4578-4019}}
\author[10,11]{Hengxing Pan \orcidlink{0000-0002-9160-391X}}
\author[8,9,7,2]{Stefano Camera \orcidlink{0000-0003-3399-3574}}
\author[12,13,2]{José Fonseca \orcidlink{0000-0003-0549-1614}}
\author[14,2]{Liantsoa F. Randrianjanahary \orcidlink{0000-0002-7600-7386}}
\author[15,16]{Gabriella De Lucia \orcidlink{0000-0002-6220-9104}}
\author[17,2]{Marta Spinelli \orcidlink{0000-0003-0148-3254}}

\affiliation[1]{Jodrell Bank Centre for Astrophysics, University of Manchester, Manchester, M13 9PL, United Kingdom}
\affiliation[2]{Department of Physics and Astronomy, University of Western Cape, Cape Town 7535, South Africa}
\affiliation[3]{Instituto de Física Teórica UAM-CSIC, Campus de Cantoblanco, 28049 Madrid, Spain}
\affiliation[4]{Institut de Recherche en Astrophysique et Plan\'etologie (IRAP), Universit\'e de Toulouse, CNRS, UPS, CNES, 14 Av. Edouard Belin, 31400 Toulouse, France}
\affiliation[5]{Universit\'e St Joseph; Faculty of Sciences, Beirut, BP-11514, Lebanon}
\affiliation[6]{Institute for Particle- and Astrophysics, ETH Z\"urich, Wolfgang-Pauli-Strasse 27, 8093 Z\"urich, Switzerland}
\affiliation[7]{INAF, Osservatorio Astrofisico di Torino, Via Osservatorio 20, 10025 Pino Torinese, Italy}
\affiliation[8]{Dipartimento di Fisica, Universit\`a degli Studi di Torino, Via P.\ Giuria 1, 10125 Torino, Italy}
\affiliation[9]{INFN, Sezione di Torino, Via P.\ Giuria 1, 10125 Torino, Italy}
\affiliation[10]{National Astronomical Observatories, Chinese Academy of Sciences, Beijing 100101, China}
\affiliation[11]{Guizhou Radio Astronomical Observatory, Guizhou University, Guiyang 550000, China}
\affiliation[12]{Instituto de Astrof\'isica e Ci\^encias do Espa\c{c}o, Universidade do Porto CAUP, 4150-762 Porto, Portugal}
\affiliation[13]{Departamento de F\'isica e Astronomia, Faculdade de Ci\^{e}ncias, Universidade do Porto, Rua do Campo Alegre 687, 4169-007 Porto, Portugal}
\affiliation[14]{Astrophysics Research Centre \& School of Mathematics, Statistics and Computer Science, University of KwaZulu-Natal, Durban, 4041, South Africa}
\affiliation[15]{INAF, Osservatorio Astronomico di Trieste, Via Tiepolo 11, 34131 Trieste, Italy}
\affiliation[16]{IFPU -- Institute for Fundamental Physics of the Universe, Via Beirut, 2, 34151 Trieste, Italy}
\affiliation[17]{Observatoire de la C\^ote d’Azur, Laboratoire J-L Lagrange, Boulevard de l'Observatoire, Nice, France}

\abstract{The 21cm line from neutral hydrogen is expected to be a ubiquitous (albeit faint) tracer of galaxies in the late Universe. With SKAO-MID, large wide-field surveys of several million H{\sc\,i}-containing galaxies will become feasible, resulting in catalogues of sufficient size to measure large-scale structure observables such as baryon acoustic oscillations and redshift-space distortions. While optical galaxy surveys over comparable areas are generally deeper, radio surveys of this kind have a number of other advantages, such as broader sampling of the halo mass function and the possibility of measuring luminosity distances via the Tully-Fisher relation. In this chapter, we provide predictions for the galaxy number counts versus redshift that will be achievable with a wide-field H{\sc\,i} galaxy survey on SKAO-MID, along with corresponding forecasts for cosmological observables. Given the substantial uncertainty in the H{\sc\,i} mass function with redshift, we bracket our predictions using a handful of different modelling methods.}

\begin{document}
\maketitle

\section{Introduction}

One of the original motivations for the SKAO project was the `Hydrogen Array' concept \citep{1991ASPC...19..428W}, an idea to detect the majority of the galaxies containing substantial reservoirs of neutral hydrogen gas out to redshifts of $z\sim 2$ and beyond. This would provide a tremendously valuable 3D map of the Universe over a large fraction of cosmic history, allowing a range of cosmological and astrophysical measurements to be carried out, from measurements of matter clustering to a detailed census of the evolution of neutral hydrogen as star formation proceeded throughout the Universe \citep{2012arXiv1212.3497E}. The necessary collecting area for such an instrument was roughly one square kilometre.

Today, optical and near-infrared spectroscopic galaxy surveys have become extremely efficient, rapidly measuring many galaxy spectra simultaneously, and therefore permitting huge samples of hundreds of millions of galaxies to be built up across survey areas of over ten thousand square degrees, and across a wide range of redshifts, even exceeding $z \sim 2$ \citep{2020MNRAS.498.2354R, 2022A&A...662A.112E, 2025arXiv250314745D}. This early historical mission of the SKA has arguably been achieved to a large extent by these other machines. Radio galaxy surveys, particularly spectroscopic ones with the 21cm line from neutral hydrogen, still have an interesting and unique role to play in modern cosmology however \citep{yahya10.1093/mnras/stv695, 2016ApJ...817...26B, 2019MNRAS.486.5124O, 2020PASA...37....7S}.

The first, and most obvious, is that neutral hydrogen gas is a crucial ingredient of star formation, and understanding how its abundance, distribution, and lifecycle change over time remains an important open question. The neutral hydrogen fraction,
\begin{equation}
\Omega_{\rm HI}(z) = \frac{\rho_{\rm HI}(z)}{\rho_{\rm crit}(z)},
\end{equation}
where $\rho_{\rm crit}(z)$ is the critical density of the Universe as a function of redshift, is poorly constrained despite numerous attempts to measure it using methods ranging from damped Lyman-$\alpha$ (DLA) surveys, 21cm intensity mapping, and resolved H{\sc\,i} galaxy surveys \citep{2017MNRAS.470..340P, 2017MNRAS.471.1788C}. Its clustering properties, particularly in relation to optical galaxies, are also relatively poorly constrained. Deeper H{\sc\,i} galaxy surveys, with much larger galaxy sample sizes, are needed to improve the statistics of these measurements at lower redshift, and provide more effective catalogues for multi-wavelength cross-correlation studies.

Secondly, the standard `concordance' cosmology model, $\Lambda$CDM, is starting to develop cracks under the weight of increasingly precise observations from a broad range of sources. When multiple observables are compared, such as various combinations of cosmic microwave background (CMB) anisotropies, supernova distance measures, galaxy clustering, and weak gravitational lensing, inconsistencies and mis-matches in the inferred cosmological parameters are being found \citep{2025RSPTA.38340022E}. This is exemplified by a multitude of independent attempts to measure the Hubble parameter, $H_0$, which is the cosmic expansion rate today. Depending on whether observables are sensitive to early- or late-time physics, the measurement seems to differ by a few km/s/Mpc, which is several times the individual measurement uncertainties \citep{2021ApJ...919...16F}. The measurements are said to be in tension, with no generally satisfactory explanation based on new physics or systematic errors yet found \citep{2021CQGra..38o3001D}. Other cosmic tensions also appear to be setting in, including tensions in measurements of the matter density fraction $\Omega_{\rm m}$ and amplitude of matter fluctuations $\sigma_8$. Perhaps even more tantalisingly, galaxy clustering measurements from the Dark Energy Spectroscopic Instrument (DESI) experiment are consistent with a dynamical dark energy field \citep{2025PhRvD.112h3515A}, and increasingly in tension with the default `cosmological constant' scenario that underpins the $\Lambda$CDM model. New observables with different properties and systematic uncertainties are therefore increasingly sought-after as a way of providing independent cross-checks on the tensions and alternative cosmological models that are being proposed.

Finally, radio instrumentation has some unique features and properties beyond optical and NIR survey telescopes. This includes insensitivity to atmospheric conditions and different angular resolution considerations (e.g. reducing effects such as galaxy blending or contamination by bright stars), full-field spectroscopy (permitting different types of blind galaxy detection and intensity mapping), insensitivity to dust extinction modelling; and general differences in the observable galaxy populations and their biases and selection functions. Beyond measuring galaxy positions and redshifts, radio surveys can also be used to measure properties such as galaxy rotation curves and line widths for example, permitting large and uniform surveys of the Tully-Fisher relation \citep[e.g.][]{2008MNRAS.391.1712M, 2016A&A...593A..39P, 2023ApJ...950...87B, 2024MNRAS.533.1550B}.

In this chapter, we provide updated predictions for the number counts and other basic properties that can be expected from a cosmological H{\sc\,i} galaxy survey. Due to the uncertainties in the H{\sc\,i} density and mass function mentioned above, we attempt to bracket the uncertainty on our predictions using a handful of different modelling approaches. We then use these predictions to forecast the constraints that should be possible on a variety of cosmological parameters from such a survey. This includes a discussion of alternative observables, such as peculiar velocities from the Tully-Fisher relation. Finally, we discuss some of the practical considerations in carrying out a sufficiently wide and deep H{\sc\,i} galaxy survey.


\section{Galaxy redshift surveys}

The 3D positions of galaxies are expected to faithfully trace the underlying dark matter distribution on large distance scales \citep{2018PhR...733....1D, 2023OJAp....6E..39A}. In regions where the galaxy number density $n_{\rm gal}$ is higher, the total matter density $\rho_m$ is expected to be correspondingly larger, i.e. we can write
\begin{equation}
\frac{n_{\rm gal}(\vec{x}) - \bar{n}_{\rm gal}}{\bar{n}_{\rm gal}} = b \left ( \frac{\rho_{\rm m}(\vec{x}) - \bar{\rho}_{\rm m}}{\bar{\rho}_{\rm m}} \right ) = b\,\delta_{\rm m}(\vec{x}),
\end{equation}
where $\delta_m$ is the matter density contrast or `overdensity', and overline denote spatial averages over large regions. The proportionality constant $b$ is the galaxy bias, and on large scales is expected to be constant (for fixed redshift).

The statistical properties of the overdensity field can be predicted from cosmological perturbation theory, which links the stochastic initial conditions set by inflation to the observed properties of the large-scale structure as a function of scale and redshift at much later times. The fundamental quantity that we are able to predict is the matter power spectrum $P(k, z)$, i.e. the variance of the matter density fluctuations,
\begin{equation}
\langle \delta_{\rm m}(\vec{k}, z)^* \delta_{\rm m}(\vec{k}^\prime, z) \rangle = (2\pi)^3 \delta^{(3)}(\vec{k} - \vec{k}^\prime) P(k, z)
\end{equation}
as a function of distance scale, represented by Fourier wavenumber $k$. The angle brackets denote an average over the statistical ensemble. The power spectrum can be predicted as a function of cosmological parameters, while measuring the fluctuations in galaxy number density over a large volume allows us to infer the power spectrum observationally. Hence, the galaxy distribution can be used to constrain cosmological models.

There are a number of caveats and subtleties to this simplified picture. First, the connection between the galaxies and the dark matter halos in which they reside must be modelled accurately. This is typically done using methods such as the `halo occupation distribution' \citep{2002ApJ...575..587B} or `sub-halo abundance matching' \citep{2006ApJ...647..201C}, which predict the number and spatial distribution of galaxies on small scales as a function of the host halo mass. Operationally, these approaches provide simple flexible models with free parameters that can be jointly fit from observations or modelled using simulations. Further complexities arise on small distance scales, where non-linear gravitational collapse and baryonic processes such as AGN feedback become important, requiring corrections to the power spectrum model. Another source of corrections is the peculiar velocities of the galaxies themselves, which introduce small distortions into the observed redshift (and therefore the inferred 3D position) of the galaxies. Observational nuances such as incompleteness and variable depth of the galaxy sample; uncertainties in galaxy redshift; and masking of contaminated or unobserved regions also require more sophisticated modelling to successfully connect the theory and observations. This is a mature field however and a range of well-tested methods exist to permit robust analyses.

\subsection{BAO and distance constraints}

Instead of making detailed predictions of the shape of the matter power spectrum, it is possible to construct observables from particular features or markers within the power spectrum, which may be less sensitive to model assumptions, e.g. about non-linear corrections and baryonic effects. The most widely-used such feature is the baryon acoustic oscillation (BAO) scale, which is a small excess in matter clustering around comoving scales of 100~Mpc \citep{2010dken.book..246B}. The BAO are formed in the early Universe, and freeze into the matter distribution shortly after the time of recombination at $z \approx 1090$. The BAO scale corresponds to the acoustic horizon at this time, and can be measured and accurately predicted from cosmic microwave background (CMB) observations. This allows the BAO scale to be used as a `standard ruler'; matching the observed scale of the BAO feature with the scale inferred from the CMB allows the distance-redshift relation to be measured. This, in turn, can be used to measure cosmological parameters such as the expansion rate, fractional matter density, and dark energy equation of state \citep{2003ApJ...598..720S}.

The BAO scale can be measured either from the Fourier-space power spectrum or, more commonly, from the {\it galaxy correlation function},
\begin{equation}
\xi_{\rm gal}(\vec{r}) \propto b^2\,\int d^3k\, P(\vec{k})\, e^{-i \vec{k}\cdot \vec{r}}.
\end{equation}
This quantity is more directly measurable from a catalogue of galaxy positions. The correlation function can be decomposed into multipoles or otherwise split into anisotropic components. The angular and redshift separations of the BAO feature correspond to the quantities
\begin{equation}
\Delta \theta_{\rm BAO} = \frac{r_{\rm BAO}}{d_A(z)};~~~~~~~ \Delta z_{\rm BAO} = \frac{H(z)}{c} r_{\rm BAO},
\label{eq:BAO_params}
\end{equation}
where $r_{\rm BAO}$ is the comoving separation of the BAO feature (inferred from the CMB) and $d_A$ and $H$ are the angular diameter distance and expansion rate respectively. By binning the galaxy samples into relatively coarse redshift bins and measuring the correlation function and the BAO feature in each, we can reconstruct the redshift-dependent $d_A(z)$ and $H(z)$ functions, which are themselves functions of the main cosmological parameters.

The BAO scale is largely insensitive to non-linearities and other effects that are difficult or complicated to model, but some care is still needed \citep{2007ApJ...664..660E}. Peculiar velocities slightly distort and shift the BAO feature, but can be partially corrected using a method called {\it velocity reconstruction}, in which linear perturbation theory is used to shift the galaxy positions back to their predicted comoving positions \citep{2007ApJ...664..660E}. The binned correlation function measurements are also quite strongly correlated as a function of separation $\vec{r}$, and so careful evaluation of their covariance matrix is also required to avoid under-estimates of the errorbars. For an extensive overview on state-of-the-art BAO modelling applied to the DESI dataset, we refer readers to \cite{10.1093/mnras/stae2090}.


\subsection{RSD constraints}

When the radial position of a galaxy is measured using its observed redshift, an additional Doppler contribution arises due to the galaxy's peculiar motion with respect to the uniform cosmic expansion. This causes a shift in its apparent position as viewed by the observer, which translates into a coherent squashing of the galaxy correlation function as a function of the angle from the line of sight. This is known as the redshift-space distortion (RSD) effect \citep{1987MNRAS.227....1K}. By measuring the degree of squashing, an estimate of the magnitude of the peculiar velocities can be made, which in turn depends on the linear growth rate of structure, $f(z)$. Dark energy and modified gravity theories can alter the relationship between the growth rate and other measures of cosmic structure, e.g. geometric measures like $d_A(z)$ and the expansion rate $H(z)$, making this a powerful observable for testing alternative gravitational theories \citep{2016ApJ...817...26B}.

A simple model for the effect of redshift-space distortions on the galaxy power spectrum is
\begin{equation}
P(k, z, \mu) = \left (b(z) + f(z) \mu^2 \right )^2\, F(z, \mu)\, P_m(k, z),
\label{eq:Pkzmu}
\end{equation}
where $\mu=\cos\theta$ is the angle between the line of sight and the centre of the galaxy sample, $b$ is the linear galaxy bias, $f$ is the linear growth rate, $F$ is a non-linear `fingers of God' (FoG) suppression term, and $P_m$ is the linear matter power spectrum. The FoG term is a non-linear counterpart to the linear RSD, and tends to result from incoherent (random) peculiar velocities on small scales, which effectively randomise the galaxy positions and hence suppress the observed clustering signal.

The linear RSD (scaled by the galaxy bias) can be extracted by taking the quadrupole moment of the anisotropic power spectrum. Implicitly, there is also a dependence on the amplitude of the power spectrum, which is proportional to the square of the linear growth factor, $D(z)$, and the amplitude parameter $\sigma_8^2$. From the anisotropic power spectrum in redshift-space, we can measure two combinations of the bias, amplitude, and linear growth rate, which we will denote as $b \sigma_8 \equiv b(z) D(z) \sigma_8(z=0)$ and $f \sigma_8 \equiv f(z) D(z) \sigma_8(z=0)$.

\subsection{Distance constraints from the power spectrum turnover}

The peak of the matter power spectrum or the turnover (TO) scale corresponds to the horizon size $r_\mathrm{H}$ at the time of matter-radiation equality. Like the BAO, it can also serve as a standard ruler, providing an independent probe and additional information on cosmological physics without making detailed predictions about the global shape of the matter power spectrum, defining $\Delta\theta_\mathrm{TO}$ and $\Delta z_\mathrm{TO}$ analogously to Eq.~\ref{eq:BAO_params}, replacing $r_\mathrm{BAO}$ with $r_\mathrm{H}$.

As the TO feature is at much larger scales than the BAO feature, large-scale structure surveys have only recently probed enough volume to detect it \citep{2025OJAp....8E..42A, 2025PhRvD.112f3553B}. Due to the scarcity of Fourier modes at the largest scales, constraints derived from the TO are inherently weaker than BAO constraints. Therefore, we will only be able to measure the angle-averaged quantity $\left(\Delta\theta_\mathrm{TO}^2\Delta z_\mathrm{TO}\right)^\frac{1}{3}$ in the foreseeable future.

Measuring the TO scale has certain advantages however. 
First, the TO standard ruler size is straightforward to compute given the ingredients of the Universe before matter-radiation equality: $r_\mathrm{H} = c\int_0^{a_\mathrm{eq}}\frac{da}{a^2 H(a)}$, where $a_\mathrm{eq}$ is the scale factor at matter-radiation equality. Second, it probes a feature set at the equality redshift $z_\mathrm{eq}\approx 3400$ and, thus, gives a handle on hypothetical models of early dark energy that may have dissipated away before the BAO redshift of $z\approx 1090$.

Finally, in a flat $\Lambda$CDM cosmology, measuring $\Delta\theta_\mathrm{BAO}$ and $\Delta z_\mathrm{BAO}$ at different redshifts yields a direct measurement of the present epoch matter density parameter $\Omega_\mathrm{m}$, but the present-epoch expansion rate $H_0$ is degenerate with $r_\mathrm{BAO}$. Assuming standard neutrino physics, $r_\mathrm{H}$ depends only on $\Omega_\mathrm{m}H_0^2$ in the same cosmological model. We can thus break degeneracies by combining uncalibrated BAO and TO measurements and obtain $\Omega_\mathrm{m}$ and $H_0$ without any external data such as the CMB for consistency checks and potentially further insights into the Hubble tension.

\section{Survey specifications}

In this section, we calculate the expected sensitivity for detecting the 21cm line from individual galaxies, based on the currently available SKA-Mid receiver and array specifications.

The point source flux density rms measured by an interferometer is given by
\begin{equation}
    S_{\rm rms} \approx \frac{2k_B T_{\rm sys}}{A_{\rm eff} \sqrt{2 \, \delta \nu \, t_p}},
\end{equation}
where $k_B$ is the Boltzmann constant, $T_{\rm sys}$ is the system temperature, $A_{\rm eff}$ is the total effective collecting area, $\delta \nu$ is the frequency channel width, and $t_p$ is the observation time per pointing. This expression has been derived in the limit of a large number of dishes, and results in a sensitivity per synthesised beam, i.e. it is in units of Jy/beam. For typical instrument specifications, this can be re-written as
\begin{equation}
\label{eqn:fluxrms}
    S_{\rm rms} = 260 \, \mu\textrm{Jy}/\textrm{beam} \left( \frac{T_{\rm sys}}{20 \textrm{\, K}}\right)
    \left( \frac{2.5 \times 10^4 \textrm{\,m}^2}{A_{\rm eff}}\right) \left( \frac{0.01 \textrm{MHz}}{\delta \nu}\right)^{1/2} \left( \frac{1 \textrm{hr}}{t_p}\right)^{1/2}.
\end{equation}
Assuming a total useful observing time for a survey $t_{\rm tot}$, divided uniformly across a survey area $S_{\rm area}$, the time per pointing can be written as
\begin{equation}
    t_p = \frac{t_{\rm tot}}{S_{\rm area} }  \left[\frac{\pi}{8} \left( \frac{1.3 \lambda }{D} \right)^2 \right],
\end{equation}
where $\lambda$ is the wavelength, and $D_{\rm dish}$ is the dish diameter. The value in the square bracket gives the area per pointing, assumed to be the same as the area of the primary beam at a given wavelength. For SKA-Mid, $t_p$ corresponds to 3.55 and 0.95 hours at the centre of Band 1 and Band 2 respectively, assuming a 5,000~deg$^2$ survey over 10,000~hours. These survey specifications were chosen to match the `Medium-Deep' survey proposed in \citet{2020PASA...37....7S}. A survey optimisation study using previous galaxy number density predictions was presented in \citet{2016ApJ...817...26B}.

\begin{figure}
    \centering
    \includegraphics[width=0.65\linewidth]{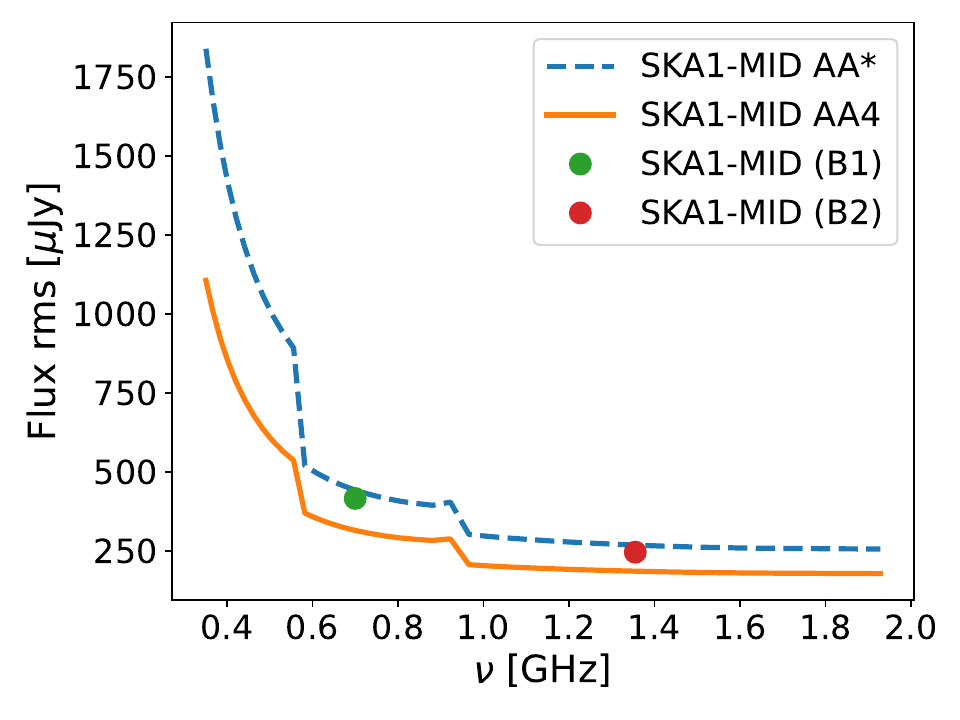}
    \caption{Expected natural sensitivity of SKA1-MID assuming 1 hour of observation time for the AA* and AA4 configurations (dashed and solid lines respectively), compared to the previous estimates from \cite{santos2015aska.confE..21S} (circles).}
    \label{fig:sensitivity}
  
\end{figure}

Using values of $A_{\rm eff} / T_{\rm sys}$ from the anticipated SKA1 science performance memo,\footnote{\url{https://www.skao.int/sites/default/files/documents/SKAO-TEL-0000818-V2_SKA1_Science_Performance.pdf}} we calculate the flux rms using Equation \ref{eqn:fluxrms} for SKA1-Mid AA* (dashed line) and AA4 (solid line) configurations with one hour of pointing and compare the values with the previous estimates from \cite{santos2015aska.confE..21S} (circles) in Figure \ref{fig:sensitivity}. A summary of the resulting survey specifications is given in Table~\ref{tbl:survey_specs}.

The H{\sc\,i} galaxy survey obtains both positions and spectroscopic redshifts simultaneously via interferometric imaging. In contrast to optical surveys such as DESI, that rely on separate imaging target selection followed by fibre spectroscopy, the interferometric data are processed into complete spectral cubes (i.e. as a function of sky pixel and frequency) that are then searched for galaxies using a source finder. Models are then fit to the candidate sources to estimate the galaxy H{\sc\,i} mass, redshift, and line width, resulting in a catalogue of sources. Because the catalogue is flux-limited, the resulting H{\sc\,i} mass sample is incomplete -- a limitation analogous to spectroscopic completeness in optical surveys.

H{\sc\,i} spectral cubes will be produced from SKA-Mid observations, with galaxies appearing as discrete H{\sc\,i} detections (S/N $\geq$ 3–5) after continuum subtraction. The best angular resolution of SKA-MID AA4 is approximately 2 arcseconds with the proper weighting scheme applied. The redshifts are derived from the line centroid with velocity resolutions of a few km s$^{-1}$, with a redshift precision of $\sim 10^{-4}$ adequate for large-scale structure science without being limited by spectral resolution.

\begin{table}[]
\begin{tabular}{|c|c|c|c|c|c|}
\hline
Configuration        & Band {[}MHz{]} & $A_{\rm eff} / T_{\rm sys}$ {[}m$^2$/K{]} & Beam FoV {[}deg$^2${]} & $t_p$ {[}hr{]} & Flux rms {[}$\mu$Jy{]} \\ \hline
\multirow{2}{*}{AA*} & 350 -- 1050     & 733.2                                     & 1.78                   & 3.55           & 235.2                  \\ \cline{2-6} 
                     & 950 -- 1760     & 1201.4                                    & 0.47                   & 0.95           & 277.8                  \\ \hline
\multirow{2}{*}{AA4} & 350 -- 1050     & 1029                                      & 1.78                   & 3.55           & 167.6                  \\ \cline{2-6} 
                     & 950 -- 1760     & 1733                                      & 0.47                   & 0.95           & 192.6                  \\ \hline
\end{tabular}
\caption{Survey specifications for a total observation time of 10,000 hours and survey area of 5,000 deg$^2$. The values for $A_{\rm eff} / T_{\rm sys}$, primary beam field of view, time per pointing $t_p$, and flux rms (per beam) are given at the central frequency of the band.}
\label{tbl:survey_specs}
\end{table}

Table~\ref{tbl:survey_specs} outlines the instrumental specifications for the SKA1-MID AA* and AA4 array configurations. These two array configurations mainly differ in their baseline distribution and compactness. This difference affects their sensitivity on different angular scales. The AA* configuration is more compact and has a high density of short baselines. This provides better surface brightness sensitivity, improving the detection of diffuse cosmic web structures and offering a wider range of H{\sc\,i} galaxy surveys by detecting fainter emission. On the other hand, the AA4 configuration is more extended and has a high density of long baselines. This provides a better angular resolution and imaging capabilities, but reduces sensitivity to large scales due to fewer short baselines \citep{Braun2024SKA1SciencePerformance}.

In this work, both configurations are considered in order to illustrate the impact of array layout on survey sensitivity and cosmological measurements. Because the galaxy detection criteria we apply only take into account the width of the spectral line and not the angular resolution (see Sect.~\ref{sec:selection}), the main impact of the different configurations is due to the difference in the corresponding values of $A_{\rm eff} / T_{\rm sys}$, as per the specifications listed in Table~\ref{tbl:survey_specs}.

We also assume a total observing time of 10,000 hours over a survey area of 5,000 square degrees, which is consistent with previous SKA1-MID H{\sc\,i} galaxy survey forecasts in the literature \citep{yahya10.1093/mnras/stv695, 2020PASA...37....7S, Spinelli:2021emp}. The relevant quantity for the thermal noise calculation is not the total survey time directly, but the effective integration time per pointing, $t_p$. The integration time per pointing is obtained by dividing the total observing time by the survey area, according to the primary beam size, to cover the full survey footprint with multiple pointings. The effective integration time per pointing can therefore be expressed in terms of the total survey time, survey area, and field of view of a single pointing. For the survey assumptions adopted here, this results in an effective integration time of approximately $t_p \approx 3.55$ hours in Band 1 and $t_p \approx 0.95$ hours in Band 2 at the band centre. This difference arises because the primary beam is smaller at higher frequencies, and therefore more pointings are required to cover the same survey area, reducing the integration time per pointing in Band 2 compared to Band 1.

\section{H{\sc\,i} galaxy simulations}
In order to properly predict cosmological observables from H{\sc\,i} galaxy survey, we need to have two important quantities: the number density ($dN/dz$) and bias with respect to the cold dark matter distribution, $b(z)$, of the detected H{\sc\,i} galaxies, both as a function of redshift and flux cut. Although they can potentially be calculated analytically, H{\sc\,i} galaxy simulations provide a more complete accounting of the physics affecting the neutral hydrogen distribution.

Following \citet{yahya10.1093/mnras/stv695}, we adopt the following simple fitting formulae:
\begin{equation} \label{eq:dNdz&bias}
\begin{split}
   \frac{dN}{dz} &= 10^{c_1} z^{c_2} \exp(-c_3 z) \\
   b(z) &= c_4 \exp(c_5 z),   
\end{split}
\end{equation}
with $c_i$ being the free parameters that best-fit the simulation data.

The number density and bias from H{\sc\,i} galaxy simulations can have large uncertainties because of the different prescriptions for cold gas evolution, ionisation, and feedback processes used in the simulations themselves \citep[e.g.][]{Pan_2020}. Other effects, such as limited resolution and box size, also affect the accuracy of the predictions. For most simulations, either on-the fly or post-processing steps are needed to resolve the hydrogen content of each galaxy \citep{Diemer_2018}.

As such, there is not yet a convergent picture of exactly what the H{\sc\,i} distribution should look like, and so any predictions we make (e.g. for the number of H{\sc\,i} galaxies that will be detected) are necessarily subject to significant uncertainty. In particular, while current simulations and models can be compared against $z \approx 0$ observations of the actual H{\sc\,i} galaxy number counts, there is relatively little information at higher redshifts. This is something that SKA-Mid itself will provide!

\subsection{Simulated H{\sc\,i} galaxy catalogues}

In this section we briefly review the three simulated catalogues used to predict the H{\sc\,i} galaxy number counts and bias for SKA-Mid.

\subsubsection{\texorpdfstring{S$^3$-SAX}{S3-SAX}}

The SKADS Simulated Skies Semi-Analytical eXtragalactic (S$^3$-SAX)\footnote{\url{http://s-cubed.physics.ox.ac.uk/s3_sax}} database consists of extragalactic radio sources enclosed in a mock lightcone that extends to $z=4$. It was based on the Millennium simulation \citep{2005Natur.435..629S}, which uses an older `WMAP-1' cosmology with a value of $\sigma_8 = 0.9$ that is somewhat higher than present estimates of $\sigma_8 \approx 0.81$. The galaxy properties are provided by the semi-analytic model of \citet{de_lucia_2007}, with further modelling to obtain H{\sc\,i} line widths, galaxy inclinations etc. We refer interested readers to \cite{Obreschkow_2009b} for more details.

In what follows, we use the best-fit parameters derived by \cite{yahya10.1093/mnras/stv695} to calculate $dN/dz$ and $b(z)$ based on the S$^3$-SAX galaxy catalogue. For convenience, we have reproduced selected values of these parameters in Table \ref{tbl:params_s3sax}.

\begin{table}[h]
\centering
\begin{tabular}{cccccc}
\hline
$S_{\rm rms}$ [$\mu$Jy] & $c_1$ & $c_2$ & $c_3$ & $c_4$  & $c_5$  \\ \hline
0             & 6.21  & 1.72  & 0.79  & 0.5874 & 9.3577 \\ \hline
1             & 6.55  & 2.02  & 3.81  & 0.4968 & 0.7206 \\ \hline
10            & 6.44  & 1.83  & 7.59  & 0.5928 & 0.8072 \\ \hline
100           & 5.63  & 1.41  & 15.49 & 0.6052 & 1.0859 \\ \hline
200           & 5.00  & 1.04  & 17.52 & --      & --      \\ \hline
\end{tabular}
\caption{Fitting parameter values for $dN/dz$ and $b(z)$ based on the S$^3$-SAX catalogue; taken from \cite{yahya10.1093/mnras/stv695}.}
\label{tbl:params_s3sax}
\end{table}

\subsubsection{GAEA} \label{sec:gaea} 

GAEA \citep{de_lucia_2014,hirschmann_2016,xie_2017,fontanot_2020,xie_2020,de_lucia_2024,fontanot_2025} is a state-of-the-art semi-analytic model (SAM) of galaxy evolution and assembly, notably presented and compared to other models in \citet{Ronconi01.2026.SKA}. The work of \citet{Obreschkow_2009b} relied on a predecessor \citep{de_lucia_2007} of GAEA to generate the S$^3$-SAX suite. In contrast, GAEA incorporates more than a decade of additional development, including up-to-date prescriptions for gas accretion, star formation, feedback, environmental quenching, and satellite galaxy evolution. In particular, the partition of cold Hydrogen in its atomic and molecular form is explicitly treated at each time step of the simulation \citep{xie_2017}, yielding more robust estimates compared to earlier methods, where it would be computed as a post-processing step on the SAM output.
H{\sc\,i} lines are associated to each galaxy in GAEA's output for the Millennium simulation, and projected on mock survey lightcones \citep{mayor_2026}.
Applying selection cuts described in \ref{sec:selection}, we obtain predictions for $\mathrm{d}N/\mathrm{d}z$ and $b(z)$, of which best fits to Equations \ref{eq:dNdz&bias} are reported in table \ref{tbl:params_gaea}.

\begin{table}[h]
\centering
\begin{tabular}{cccccc}
\hline
$S_{\rm rms}$ [$\mu$Jy] & $c_1$ & $c_2$ & $c_3$ & $c_4$  & $c_5$  \\ \hline
0             & 6.370  & 2.117  & 2.609  & 0.738 & 0.297 \\ \hline
1             & 6.339  & 2.102  & 3.091  & 0.738 & 0.300 \\ \hline
10            & 6.538  & 2.208  & 6.432  & 0.723 & 0.442 \\ \hline
100           & 6.849  & 2.336  & 17.895 & 0.710 & 0.923 \\ \hline
200           & 7.170  & 2.492  & 26.385 & 0.716 & 1.086 \\ \hline
\end{tabular}
\caption{Fitting parameter values for $dN/dz$ and $b(z)$ based on the GAEA catalogue.}
\label{tbl:params_gaea}
\end{table}

\subsubsection{IllustrisTNG}\label{sec:TNG}
The IllustrisTNG\footnote{\url{https://www.tng-project.org/}} project is a suite of state-of-the-art cosmological magnetohydrodynamical simulations that aims to realistically capture a wide range of the physical processes driving the formation and evolution of galaxies in the Universe \citep{tng2018MNRAS.475..624N, tng2018MNRAS.475..648P, tng2018MNRAS.475..676S, tng2018MNRAS.477.1206N, tng2018MNRAS.480.5113M, tng2019ComAC...6....2N}. It uses the Planck 2015 cosmology \citep{refId0} and there are three physical box sizes: 35, 75, and 205 Mpc/$h$ that are referred to TNG50, TNG100, and TNG300 respectively. In addition to having complete halo catalogues from these simulation box sizes, the IllustrisTNG project also has various supplementary data post-processed from the original catalogue, which includes the H\,\textsc{i} and H$_2$ content of the galaxies \citep{diemer2019tng}. We have chosen to use the data from TNG100 because of its higher resolution, allowing it to resolve more of the low-mass halos that are expected to contain a significant fraction of the neutral hydrogen. Note that this smaller volume may not be large enough to give fully robust estimates of the galaxy bias however.

We first outline the formulae used to convert the IllustrisTNG output to the observable quantities, before detailing the steps taken to calculate the H{\sc\,i} galaxy comoving density and bias.
We begin by converting the atomic H{\sc\,i} mass, $M_{\rm HI}$, to the intrinsic H{\sc\,i} line width, $W_{\rm e}$, following \cite{duffy2012askap}, where
\begin{equation}
    \frac{W_{\rm e}}{420 \textrm{\, km s}^{-1}} = \left(\frac{ M_{\rm HI}}{10^{10} \textrm{\, M}_\odot}\right)^{0.3}.
    \label{eqn:mass_width}
\end{equation}
We also follow the same method in the paper whereby a value of inclination angle $\theta$ is assigned to each H{\sc\,i} galaxy following a random uniform distribution in the cosine of the angle, i.e. cos($\theta) \sim \it{U}(0,1)$.
The line width of a galaxy with inclination and width $(\theta$, $W_\theta)$ can then be calculated using the Tully-Fouque rotation scheme \citep{1985ApJS...58...67T},
\begin{equation}
    (W_{\rm e} \textrm{ sin}(\theta))^2 = W_\theta^2 + V_{\rm o}^2 - W_\theta W \left[1 - \exp -\left(\frac{W_\theta}{V_{\rm c}}\right)^2\right] - 2 V_{\rm o}^2 \exp-\left(\frac{W_\theta}{V_{\rm c}}\right)^2,
    \label{eqn:tullyfouque}
\end{equation}
where $V_{\rm c}=120$ km s$^{-1}$, $V_{\rm o} \approx 20$ km s$^{-1}$, and $W \approx 38$ km s$^{-1}$.

Under the assumption that the H{\sc\,i} galaxies are optically thin, the observed-frame, velocity-integrated flux $S^{V_{\rm obs}}$ can be calculated from $M_{\rm HI}$ based on the relation
\begin{equation}
    M_{\rm HI} \simeq \frac{2.35 \times 10^5\, h^{-2} {\rm M}_\odot}{(1+z)^2} \left( \frac{D_L}{h^{-1} \rm{\,Mpc}} \right)^2 \left(\frac{S^{V_{\rm obs}}}{\rm{Jy \, km \, s}^{-1}} \right),
\end{equation}
 where $h$ is the dimensionless Hubble constant and $D_L$ is the luminosity distance \citep{Meyer_Robotham_Obreschkow_Westmeier_Duffy_Staveley-Smith_2017}. To obtain $S$, one can then divide $S^{V_{\rm obs}}$ by $W_\theta$.

Because the H\,\textsc{i} simulation data are only available at snapshot redshifts $z = [0, 0.5, 1.0, 1.5, 2.0]$, we choose to interpolate the per-snapshot data instead of constructing a lightcone. We first bin the data according to $z$, $M_{\rm HI}$, and $W_\theta$. Next, we divide the number count by the simulation volume to find the number count per volume as a function of redshift, $dN/dV(z)$, before interpolating between $z=0.01$ and $z=2$.

We then convert the quantities to $S$, apply the two selection cuts detailed in the next section, and sum the bins of $dN/dV(z)$. Finally, to obtain the proper comoving number density $dN/dz$,  we multiply them by
\begin{equation}
    \frac{dV}{dz d\Omega} \, \textrm{[Mpc}^3 \textrm{deg}^{-2} \textrm{ per unit \,} z] = \frac{c r(z)^2}{H(z)} \left(\frac{\pi}{180}\right)^2,
\end{equation}
where $c$ is the speed of light, $r(z)$ is the comoving radius at $z$, and $H(z)$ is the Hubble parameter. 

\label{sec:bz}
Similarly, to calculate the H\,\textsc{i} galaxy bias, we bin the data according to $z$, $M_{\rm HI}$, $W_\theta$, and additionally, the dark matter halo mass, $M_h$. After obtaining $S$ from $M_{\rm HI}$ and $W_\theta$, we apply the two selection cuts and calculate the H{\sc\,i} galaxy bias, $b(z, S_{\rm rms})$, following
\begin{equation}
    b(z, S_{\rm rms}) \approx \sum_i b(z, M_h^i) \frac{N(M_h^i)}{N_{\rm tot}},
\end{equation}
where $b(z, M_h^i)$, $N(M_h^i)$, and $N_{\rm tot}$ are the Sheth-Tormen halo bias for mass $M_h^i$, number of dark matter halo with mass $M_h^i$ passing the selection cuts for a particular $S_{\rm rms}$, and total number of halos passing the selection cuts respectively
\citep{yahya10.1093/mnras/stv695}.

\subsection{Galaxy detection criteria} \label{sec:selection}

To explore the range of plausible values of $dN/dz$ and $b(z)$ and attempt to get a handle on this modelling uncertainty, we have compared predictions across three simulations, which have been briefly described in the previous section. The specific modelling steps needed to generate mock galaxy catalogues with H{\sc\,i} masses, total fluxes, and line-widths is quite involved in each of these cases, and we refer the reader to the simulation description papers for full details.

For all three simulations, we applied the following selection cuts, following \cite{yahya10.1093/mnras/stv695}:
\begin{itemize}
    \item Observed H{\sc\,i} line width $W_\theta > 2 \delta V$;
    \item Total H{\sc\,i} flux $S > N_{\rm cut} \times S_{\rm rms} / \sqrt{W_\theta / \delta V}$,
\end{itemize}
where $N_{\rm cut}=5$, and the frequency resolution is taken to be 10 kHz, corresponding to a velocity width of $\delta V= 2.1 (1+z)$ km s$^{-1}$.

\begin{figure}
    \centering
    \includegraphics[width=1\linewidth]{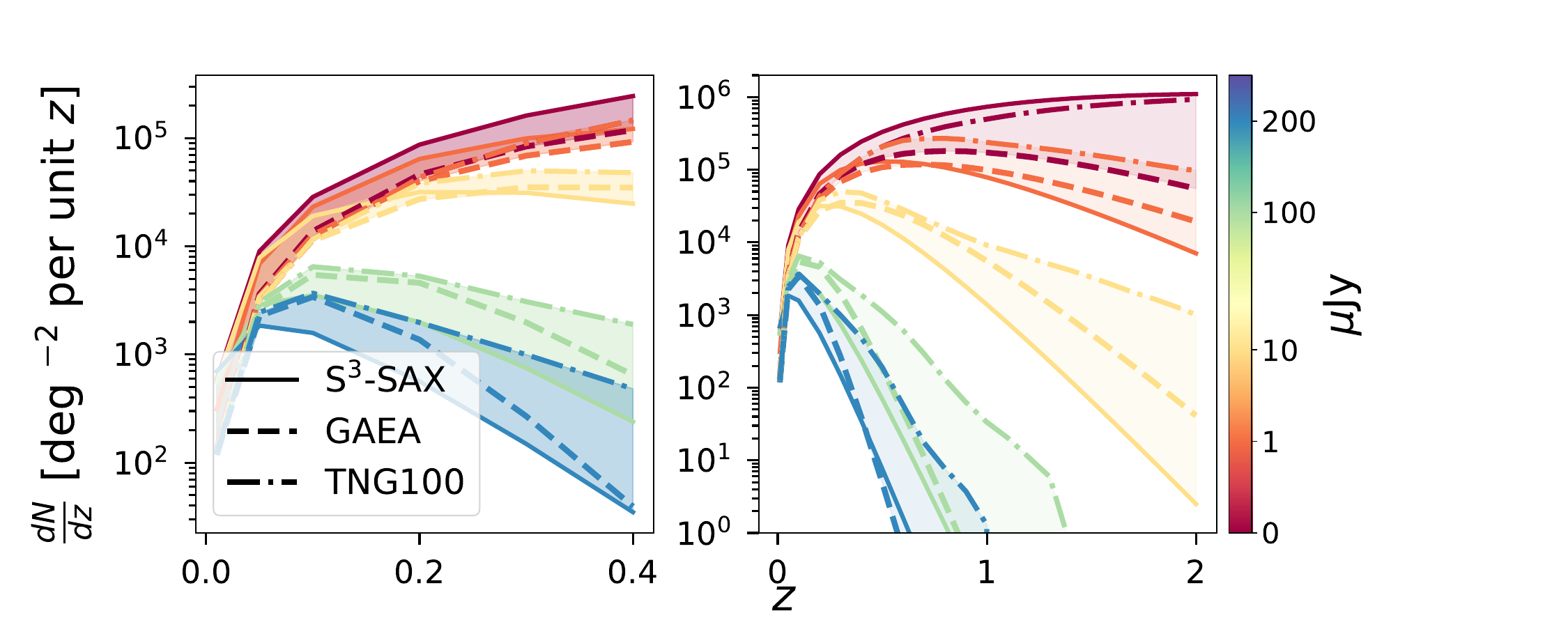}
    \caption{The angular number density with various constant flux cuts for the S$^3$-SAX (solid), GAEA (dashed), and TNG100 (dash-dot) H{\sc\,i} simulations. The left panel is a zoom-in of the $z\leq0.4$ range of the right panel. The colours denote the fixed flux rms thresholds that are labelled on the colour bar.}
    \label{fig:dndz}
  
\end{figure}

\begin{figure}
    \centering
    \includegraphics[width=1\linewidth]{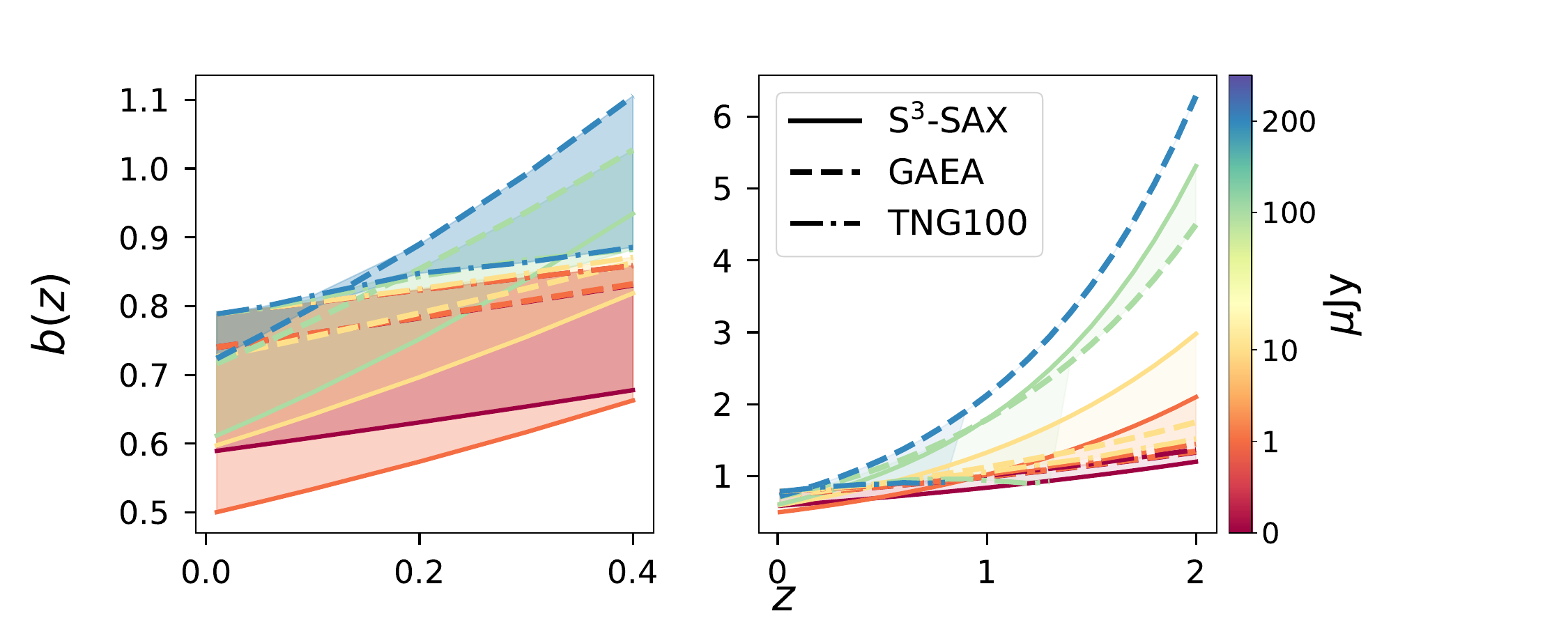}
    \caption{The H{\sc\,i} bias with various flux cuts for S$^3$-SAX (solid), GAEA (dash), and TNG100 (dash dot) H{\sc\,i} simulation. The left panel is a zoom-in of the $z\leq0.4$ range of the right panel.}
    \label{fig:bz}
\end{figure}

Predictions for $dN/dz$ and $b(z)$ for each simulation are shown in Figures \ref{fig:dndz} and \ref{fig:bz} respectively, with the colours corresponding to a selection of fixed flux rms thresholds.

It can be seen that S$^3$-SAX generally predicts the highest number density at the $0~\mu$Jy threshold, although the predictions from IllustrisTNG are quite similar, despite the very different simulation methodology. The GAEA number counts are systematically lower at this threshold, by about a factor of $2-3$ at low redshift, and over an order of magnitude at $z \gtrsim 1.5$.

The situation is reversed at larger flux rms threshold values, with the S$^3$-SAX values being lower at all but the lowest redshifts. Instead, IllustrisTNG has the larger number counts for $10~\mu$Jy and above, except for at $z \lesssim 0.05-0.1$. The GAEA predictions are mostly intermediate between the other two, although in some places approach the values of one of the other simulations. The discrepancy between predictions can run into multiple orders of magnitude for the higher redshifts, although this only happens quite far beyond the peak of each curve, where $dN/dz$ has dropped by an order of magnitude or so. The peaks themselves show reasonably good agreement between simulations, to within a factor of roughly two.

The bias, shown in Fig.~\ref{fig:bz}, exhibits a broad range of values. While there is at most only a factor of $\sim 2$ difference between the simulations at low redshift, this is quite a large discrepancy for the bias. Looking at higher redshift, H{\sc\,i} galaxies become harder to detect, and only the rarer, brighter ones should be detectable. Since these will tend to reside in the highest mass dark matter halos, they are likely to have large galaxy biases, which is indeed what we see above $z \sim 1$.

The predictions of each simulation are therefore mostly within a factor of two of one another; this is a substantial theoretical uncertainty, and the results presented below should be interpreted in this light. The fact that this discrepancy is not larger -- like the multi-orders of magnitude discrepancies seen at the highest redshifts -- is a positive outcome however. In other words, we should be able to trust the predicted H{\sc\,i} galaxy number counts and bias to within a factor of about two at the redshifts where most galaxies are expected to be detected (unless {\it all} of the simulations seriously diverge from reality).

\section{Cosmological performance}

In this section, we present basic Fisher matrix forecasts to predict how well SKA-Mid H{\sc\,i} galaxy surveys will be able to measure a selection of cosmological observables and parameters. The number density and bias predictions are now made in a series of redshift bins of width $\Delta z = 0.1$, with the flux density threshold $S_{\rm rms}$ calculated based on Equation \ref{eqn:fluxrms} now varying with frequency (evaluated at the frequency of the centre redshift in each bin).

\subsection{Forecast from BAO and RSD probes}


We use the \texttt{RadioFisher}\footnote{\url{https://github.com/philbull/RadioFisher}} Fisher forecast code \citep{Bull_2015} to predict how well a 5,000 deg$^2$ survey over 10,000 hours with SKA-Mid (AA* and AA4 configurations) will be able to measure the expansion rate $H(z)$ and angular diameter distance $d_A(z)$. The galaxy bias and linear growth rate in each redshift bin are marginalised, as is the non-linear scale Fingers of God effect that appears in the expression for the RSDs, meaning that a basic accounting for systematic effects has been included. Fisher forecasts are otherwise quite idealised however, and so will tend to be optimistic compared with analyses of real data. In particular, it should be noted that we are assuming a `full shape' analysis of the redshift-space power spectrum, in which not only the BAO and RSD features, but also the broadband shape and Alcock-Paczynski terms, are used to constrain the radial and transverse scales in each redshift bin \citep[for details, see][]{Bull_2015}. We also neglect the effect of BAO reconstruction, which can substantially improve the sharpness of the BAO feature \citep{2007ApJ...664..675E}.

\begin{figure}
    \centering
    \includegraphics[width=0.49\linewidth]{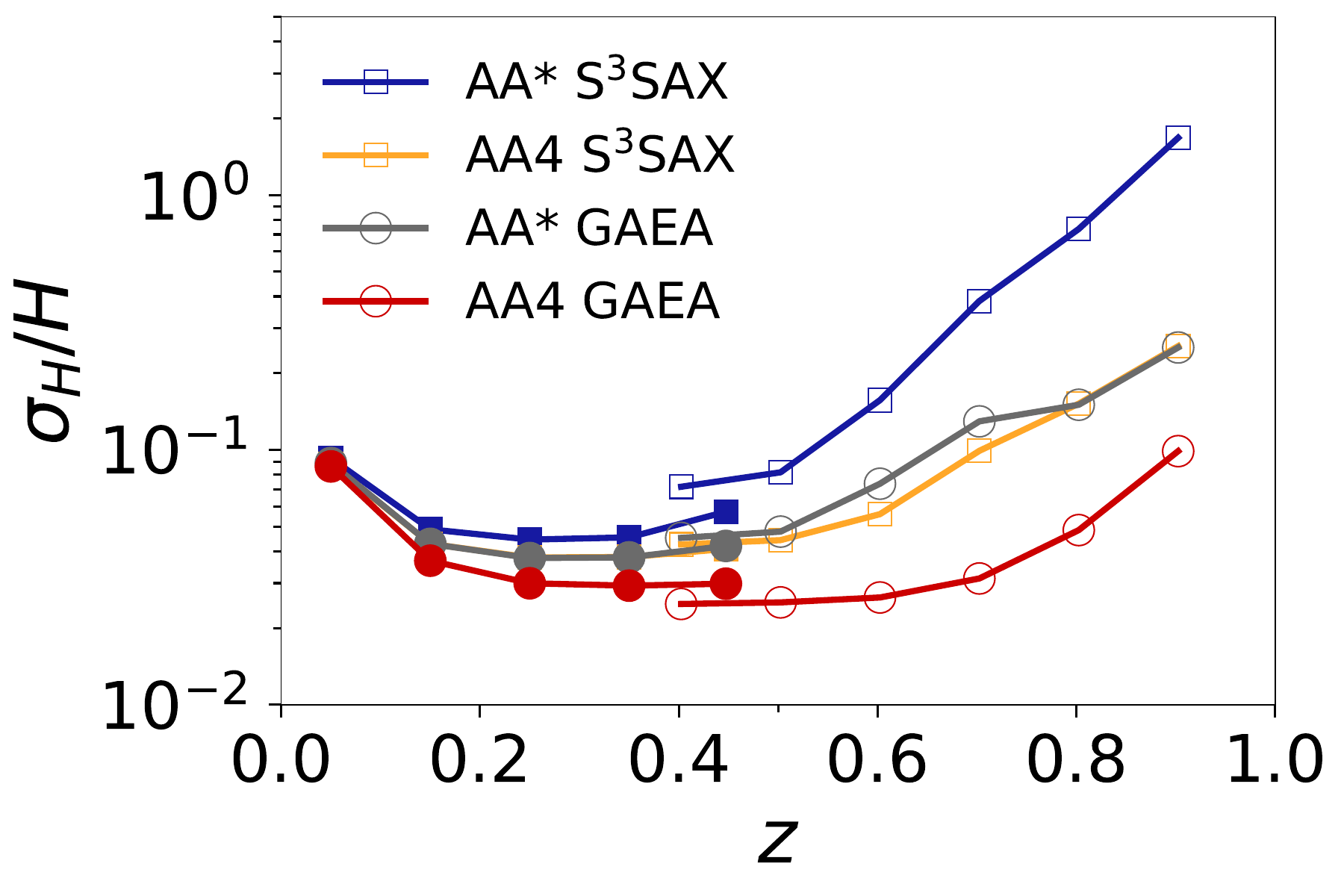}
     \includegraphics[width=0.49\linewidth]{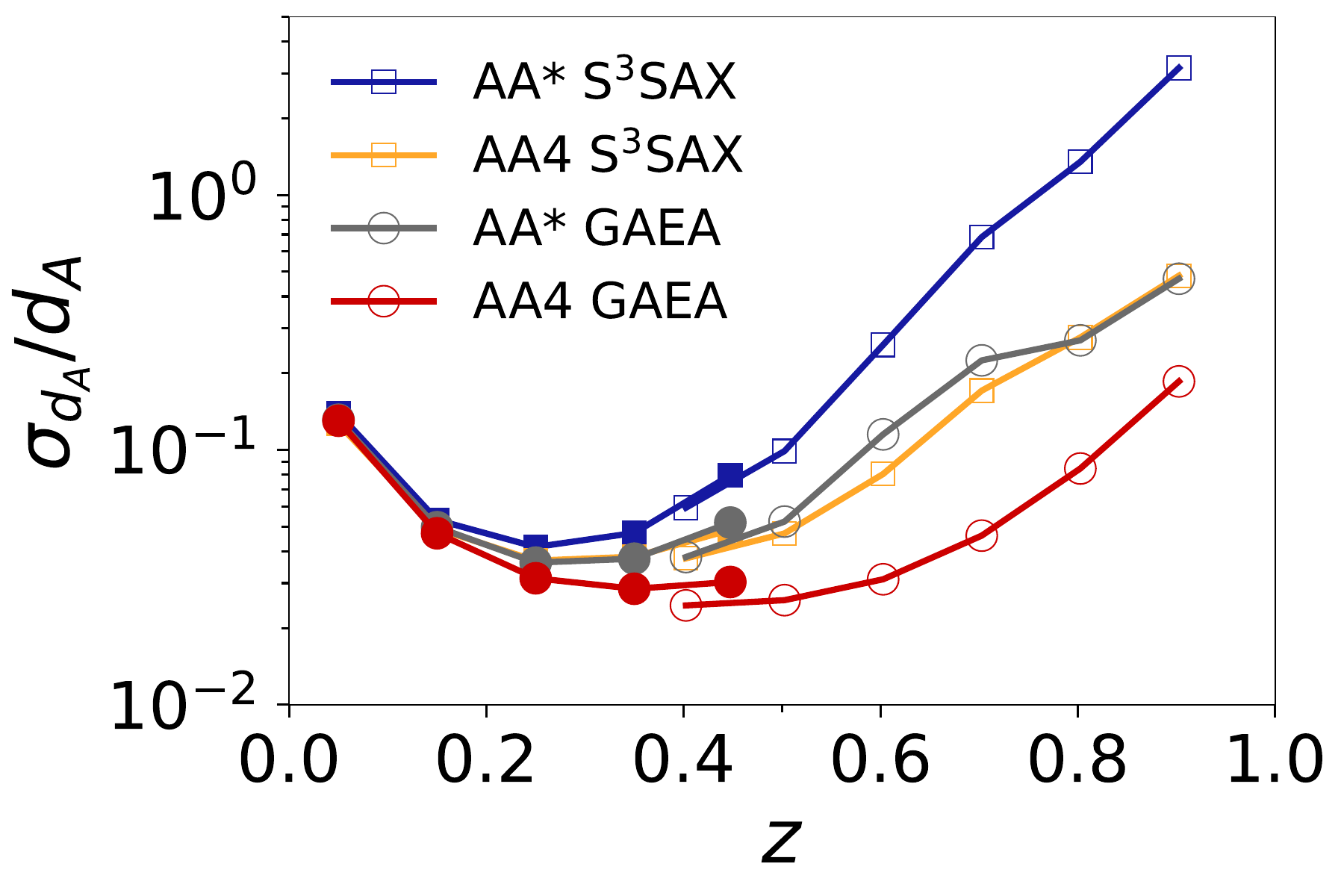}
    \caption{Forecast fractional errors on (left) the expansion rate, $H(z)$, and (right) angular diameter distance, $d_A(z)$, for the AA* and AA4 survey parameters, based on predicted number densities from S$^3$-SAX and GAEA.}
    \label{fig:hz_da}
\end{figure}

Fig.~\ref{fig:hz_da} shows the expansion rate and angular diameter distance forecasts for survey number density predictions made with the S$^3$-SAX and GAEA simulations, which bracket the range of number densities that we expect to find. The lower-$z$ lines with filled markers are for Band~2, while the higher-$z$ with unfilled ones are for Band~1. For both array configurations and bands, the constraints are better than 10\% from $z \approx 0.1 - 0.5$, reaching $2-3\%$ for $H(z)$ at $z \approx 0.4$ on both bands, for the GAEA AA4 predictions. These degrade by a factor of two or so if the simulation is changed to S$^3$-SAX and/or the AA* array configuration is used. The numbers are broadly similar for $d_A(z)$.

The predictions from the two simulations vary the most for the Band~1 points, with almost a factor of 10 difference at $z \approx 0.9$, although the constraints at this redshift are very weak. At intermediate redshifts, the difference is less pronounced. There is very good agreement between both simulations and array configurations in the lowest redshift bin. The degradation in the constraints at the lowest $z$ likely indicates that they are sample variance limited at $z \lesssim 0.15$; a larger survey area would be needed to improve these.

\begin{figure*}
    \centering
    \includegraphics[width=0.49\linewidth]{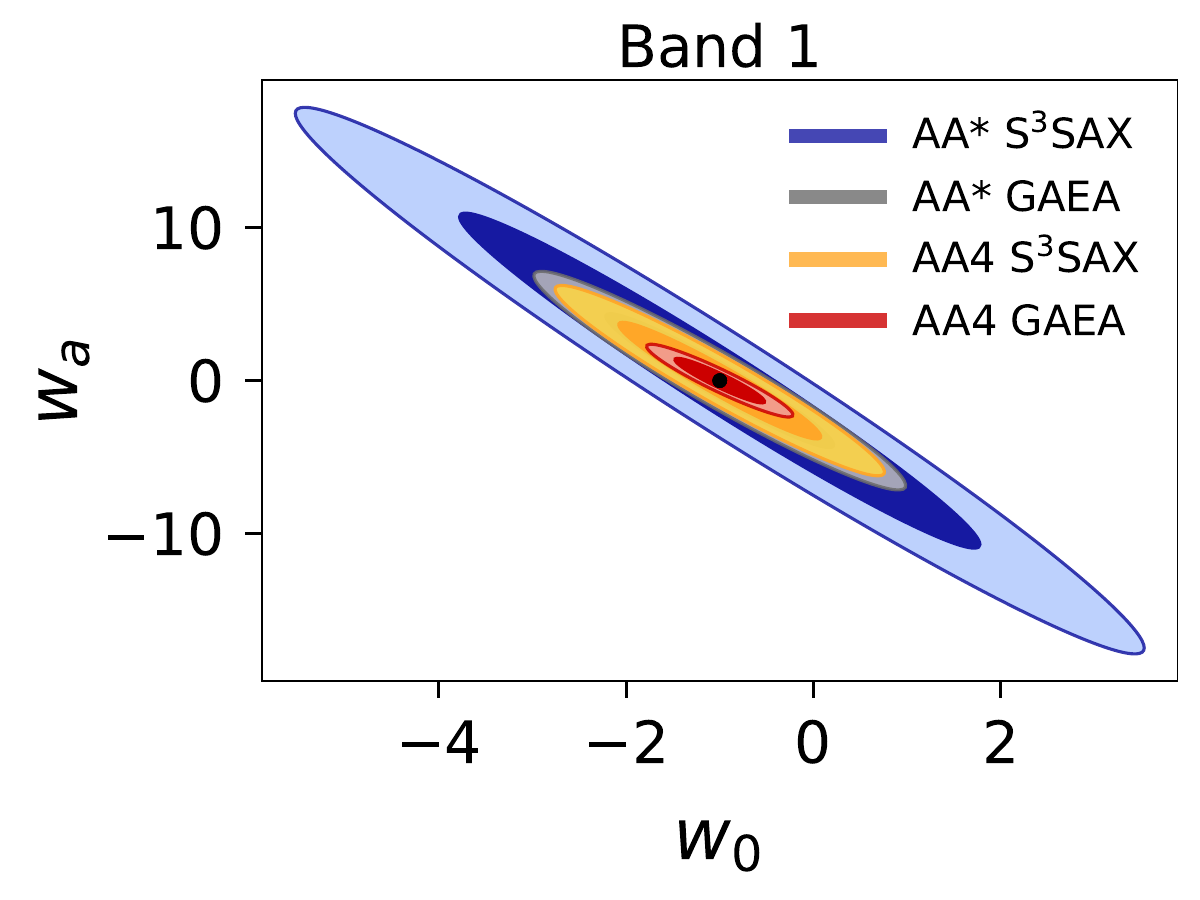}
    \includegraphics[width=0.49\linewidth]{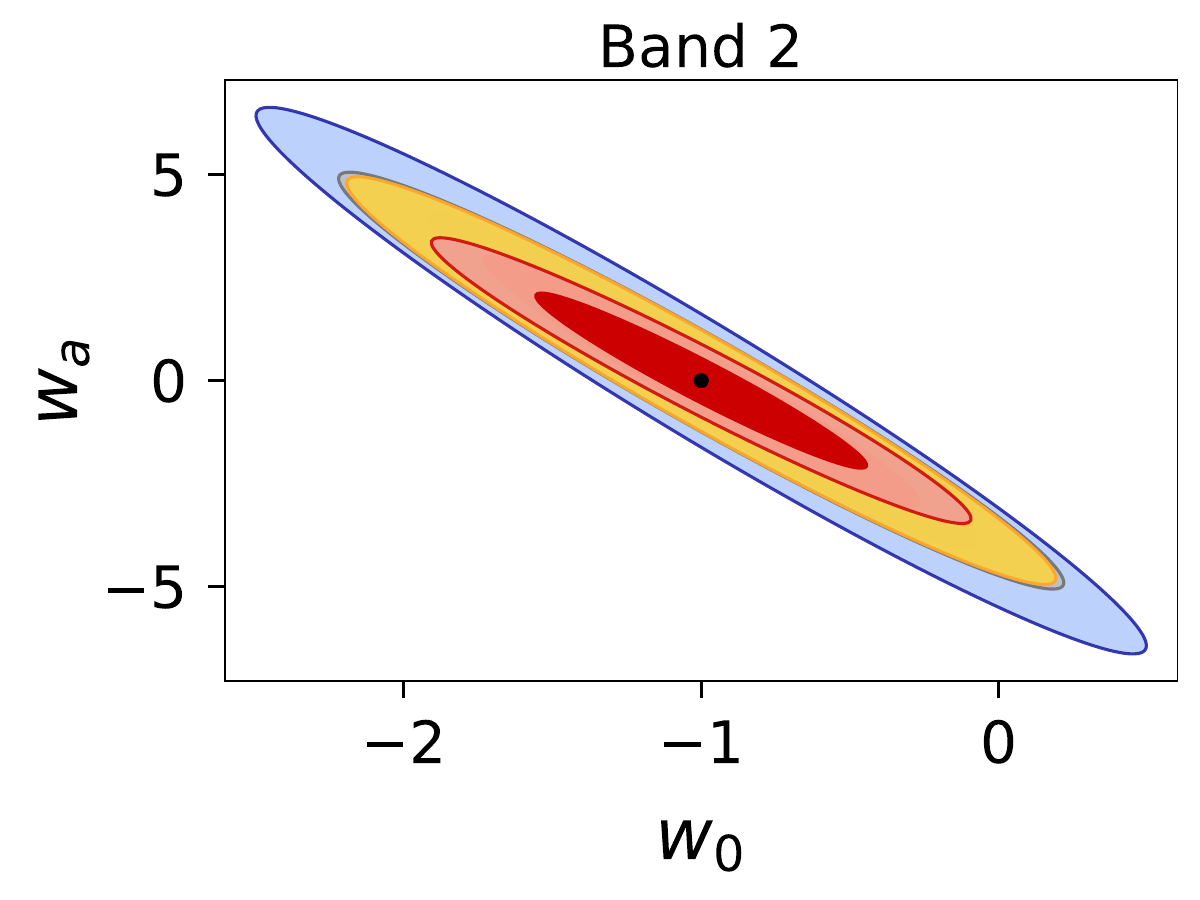}
    \caption{Forecast 1- and 2-$\sigma$ contours on the dark energy equation of state parameters $w_0,w_a$ from S$^3$-SAX and GAEA for SKA1-MID Band 1 (left) and Band 2 (right). Note the close overlap of the yellow and grey contours.}
    \label{fig:waw0}
  
\end{figure*}

Fig.~\ref{fig:waw0} shows the result of projecting the forecasts for $H(z)$ and $d_A(z)$ onto a set of cosmological parameters, for the set of 5 standard parameters, plus the dark energy equation of state parameters $w_0$ and $w_a$ (assuming a flat universe). A (Gaussianised) Planck CMB prior has also been included \citep{refId0}. This is done for both Band 1 and Band 2, for the AA4 and AA* configurations, and both the S$^3$-SAX and GAEA number count predictions. The predicted values are not competitive with those obtained by contemporary spectroscopic galaxy surveys such as DESI, but would be significantly improved if combined with (e.g.) BOSS or DESI datasets. We have not attempted to include other probes except for CMB in our forecasts here however. A summary of the forecasted constraints, along with the number of H{\sc\,i} galaxies over 5,000 deg$^2$ ($N_{\rm gal}$) and $b(z)$ are provided in Tables \ref{tbl:s3sax_forecast}, \ref{tbl:gaea_forecast} and \ref{tbl:waw0}.

\begin{table}[]
\centering
\begin{tabular}{ccc|cc|cc|}
\cline{4-7}
                                   &                                          & \textbf{}    & \multicolumn{2}{c|}{\textbf{AA*}}                                         & \multicolumn{2}{c|}{\textbf{AA4}}                                         \\ \hline
\multicolumn{1}{|c|}{$\mathbf{z}$} & \multicolumn{1}{c|}{$\mathbf{N_{\rm gal}}$} & $\mathbf{b}$ & \multicolumn{1}{c|}{$\mathbf{\sigma_H/ H}$} & $\mathbf{\sigma_{d_A}/d_A}$ & \multicolumn{1}{c|}{$\mathbf{\sigma_H/ H}$} & $\mathbf{\sigma_{d_A}/d_A}$ \\ \hline
\multicolumn{1}{|c|}{0.05}         & \multicolumn{1}{c|}{2.31 $\times 10^9$}                    & ---            & \multicolumn{1}{c|}{0.093}                  & 0.139                       & \multicolumn{1}{c|}{0.088}                  & 0.126                       \\ \hline
\multicolumn{1}{|c|}{0.15}         & \multicolumn{1}{c|}{2.51$\times 10^9$}                    & ---            & \multicolumn{1}{c|}{0.049}                  & 0.053                       & \multicolumn{1}{c|}{0.043}                  & 0.049                       \\ \hline
\multicolumn{1}{|c|}{0.25}         & \multicolumn{1}{c|}{7.41$\times 10^8$}                    & ---            & \multicolumn{1}{c|}{0.044}                  & 0.042                       & \multicolumn{1}{c|}{0.038}                  & 0.037                       \\ \hline
\multicolumn{1}{|c|}{0.35}         & \multicolumn{1}{c|}{1.82$\times 10^8$}                    & ---            & \multicolumn{1}{c|}{0.045}                  & 0.047                       & \multicolumn{1}{c|}{0.038}                  & 0.038                       \\ \hline
\multicolumn{1}{|c|}{0.5}          & \multicolumn{1}{c|}{1.91$\times 10^7$}                    & ---            & \multicolumn{1}{c|}{0.082}              & 0.099                   & \multicolumn{1}{c|}{0.044}              & 0.047                   \\ \hline
\multicolumn{1}{|c|}{0.6}          & \multicolumn{1}{c|}{4.00$\times 10^6$}                    & ---            & \multicolumn{1}{c|}{0.157}              & 0.258                   & \multicolumn{1}{c|}{0.056}              & 0.080                   \\ \hline
\multicolumn{1}{|c|}{0.7}          & \multicolumn{1}{c|}{8.14$\times 10^5$}                   & ---            & \multicolumn{1}{c|}{0.385}              & 0.683                   & \multicolumn{1}{c|}{0.099}              & 0.171                   \\ \hline
\multicolumn{1}{|c|}{0.8}          & \multicolumn{1}{c|}{1.62$\times 10^5$}                    & ---            & \multicolumn{1}{c|}{0.736}              & 1.349                   & \multicolumn{1}{c|}{0.152}              & 0.276                   \\ \hline
\multicolumn{1}{|c|}{0.9}          & \multicolumn{1}{c|}{3.18$\times 10^4$}                    & ---            & \multicolumn{1}{c|}{1.676}              & 3.151                   & \multicolumn{1}{c|}{0.255}              & 0.482                   \\ \hline

\end{tabular}

\caption{\label{tbl:s3sax_forecast} The forecasted constraints for $H(z)$ and $d_A(z)$, along with the values of $N_{\rm gal}$ and $b(z)$ per redshift bin for the S$^3$-SAX simulation.}
\end{table}

\begin{table}[]
\centering
\begin{tabular}{ccc|cc|cc|}
\cline{4-7}
                                   &                               &              & \multicolumn{2}{c|}{\textbf{AA*}}                                & \multicolumn{2}{c|}{\textbf{AA4}}                       \\ \hline
\multicolumn{1}{|c|}{$\mathbf{z}$} & \multicolumn{1}{c|}{$\mathbf{N_{\rm gal}}$} & $\mathbf{b}$ & \multicolumn{1}{c|}{$\mathbf{\sigma_H/ H}$} & $\mathbf{\sigma_{d_A}/d_A}$ & \multicolumn{1}{c|}{$\mathbf{\sigma_H/ H}$} & $\mathbf{\sigma_{d_A}/d_A}$ \\ \hline
\multicolumn{1}{|c|}{0.05}         & \multicolumn{1}{c|}{2.63$\times 10^9$}                    & 0.761        & \multicolumn{1}{c|}{0.089}         & 0.131                       & \multicolumn{1}{c|}{0.086}         & 0.130              \\ \hline
\multicolumn{1}{|c|}{0.15}         & \multicolumn{1}{c|}{4.39$\times 10^9$}                    & 0.854        & \multicolumn{1}{c|}{0.043}         & 0.050                       & \multicolumn{1}{c|}{0.037}         & 0.047              \\ \hline
\multicolumn{1}{|c|}{0.25}         & \multicolumn{1}{c|}{8.24$\times 10^8$}                    & 0.958        & \multicolumn{1}{c|}{0.038}         & 0.036                       & \multicolumn{1}{c|}{0.030}         & 0.031              \\ \hline
\multicolumn{1}{|c|}{0.35}         & \multicolumn{1}{c|}{9.93$\times 10^7$}                    & 1.074        & \multicolumn{1}{c|}{0.038}         & 0.037                       & \multicolumn{1}{c|}{0.029}         & 0.029              \\ \hline
\multicolumn{1}{|c|}{0.5}          & \multicolumn{1}{c|}{2.85$\times 10^6$}                    & 1.275        & \multicolumn{1}{c|}{0.048}     & 0.052                   & \multicolumn{1}{c|}{0.025}     & 0.026          \\ \hline
\multicolumn{1}{|c|}{0.6}          & \multicolumn{1}{c|}{2.33$\times 10^5$}                    & 1.430        & \multicolumn{1}{c|}{0.074}     & 0.115                   & \multicolumn{1}{c|}{0.026}     & 0.031          \\ \hline
\multicolumn{1}{|c|}{0.7}          & \multicolumn{1}{c|}{1.77$\times 10^4$}                    & 1.604        & \multicolumn{1}{c|}{0.129}     & 0.225                   & \multicolumn{1}{c|}{0.031}     & 0.046          \\ \hline
\multicolumn{1}{|c|}{0.8}          & \multicolumn{1}{c|}{1.27$\times 10^3$}                    & 1.798        & \multicolumn{1}{c|}{0.150}     & 0.269                   & \multicolumn{1}{c|}{0.048}     & 0.085          \\ \hline
\multicolumn{1}{|c|}{0.9}          & \multicolumn{1}{c|}{8.79$\times 10^1$}                    & 2.017        & \multicolumn{1}{c|}{0.252}     & 0.469                   & \multicolumn{1}{c|}{0.099}     & 0.186          \\ \hline
\end{tabular}

\caption{\label{tbl:gaea_forecast} The forecasted constraints for $H(z)$ and $d_A(z)$, along with the values of $N_{\rm gal}$ and $b(z)$ per redshift bin for the GAEA simulation.}
\end{table}

\begin{table}[]
\centering 
\begin{tabular}{c|ccc|ccc|}
\cline{2-7}
\textbf{}                                    & \multicolumn{3}{c|}{\textbf{S$^3$-SAX}}                                                                    & \multicolumn{3}{c|}{\textbf{GAEA}}                                                                         \\ \hline
\multicolumn{1}{|c|}{\textbf{Configuration}} & \multicolumn{1}{c|}{$\mathbf{\sigma_{w_0}}$} & \multicolumn{1}{c|}{$\mathbf{\sigma_{w_a}}$} & \textbf{FOM} & \multicolumn{1}{c|}{$\mathbf{\sigma_{w_0}}$} & \multicolumn{1}{c|}{$\mathbf{\sigma_{w_a}}$} & \textbf{FOM} \\ \hline
\multicolumn{1}{|c|}{SKA AA* Band 1}         & \multicolumn{1}{c|}{1.83}                  & \multicolumn{1}{c|}{7.20}                  & 0.36       & \multicolumn{1}{c|}{0.80}                  & \multicolumn{1}{c|}{2.88}                  & 1.55       \\ \hline
\multicolumn{1}{|c|}{SKA AA* Band 2}         & \multicolumn{1}{c|}{0.60}                  & \multicolumn{1}{c|}{2.68}                  & 2.54       & \multicolumn{1}{c|}{0.49}                  & \multicolumn{1}{c|}{2.04}                  & 4.06       \\ \hline
\multicolumn{1}{|c|}{SKA AA4 Band 1}         & \multicolumn{1}{c|}{0.71}                  & \multicolumn{1}{c|}{2.51}                  & 2.01       & \multicolumn{1}{c|}{0.32}                  & \multicolumn{1}{c|}{0.96}                  & 9.77       \\ \hline
\multicolumn{1}{|c|}{SKA AA4 Band 2}         & \multicolumn{1}{c|}{0.27}                  & \multicolumn{1}{c|}{1.85}                  & 2.52       & \multicolumn{1}{c|}{0.37}                  & \multicolumn{1}{c|}{1.40}                  & 7.65       \\ \hline
\end{tabular}
\caption{\label{tbl:waw0} The forecasted 1$\sigma$ marginal errors and dark energy Figure of Merit (FOM) for the different simulation and instrument configuration.}
\end{table}

\subsubsection{Robustness of forecasts}

Different approaches to nuisance parameter marginalisation and modelling of noise and cosmological functions can result in substantially different forecasts for what should otherwise be the same observables. To probe this, we also perform an alternative set of forecasts using an alternative Fisher matrix approach.

\begin{figure*}
    \centering
    \includegraphics[width=0.49\linewidth]{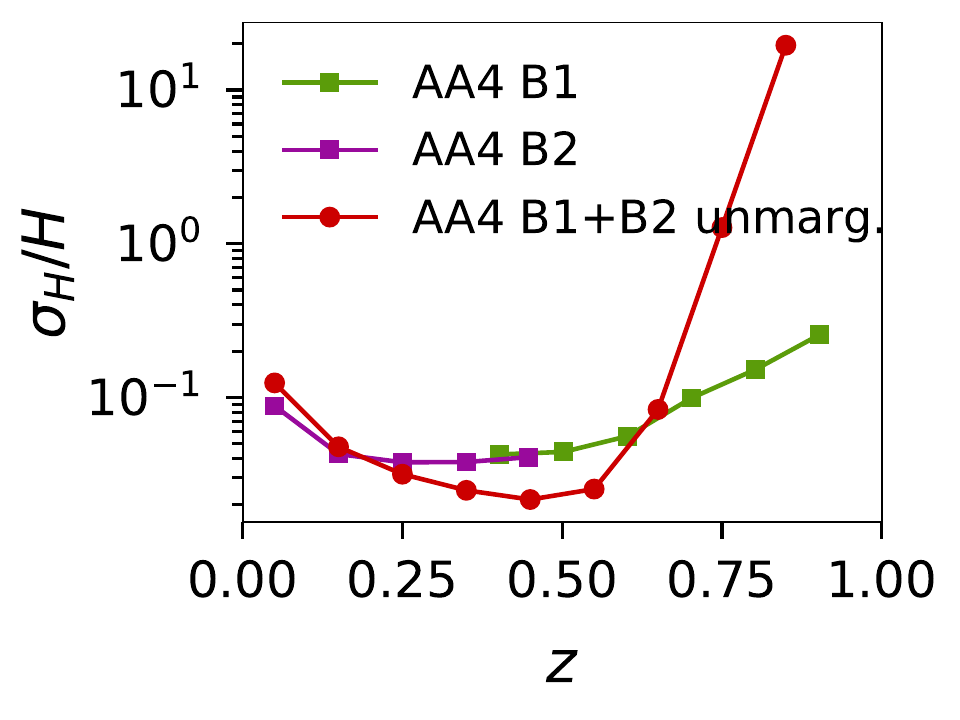}
    \includegraphics[width=0.49\linewidth]{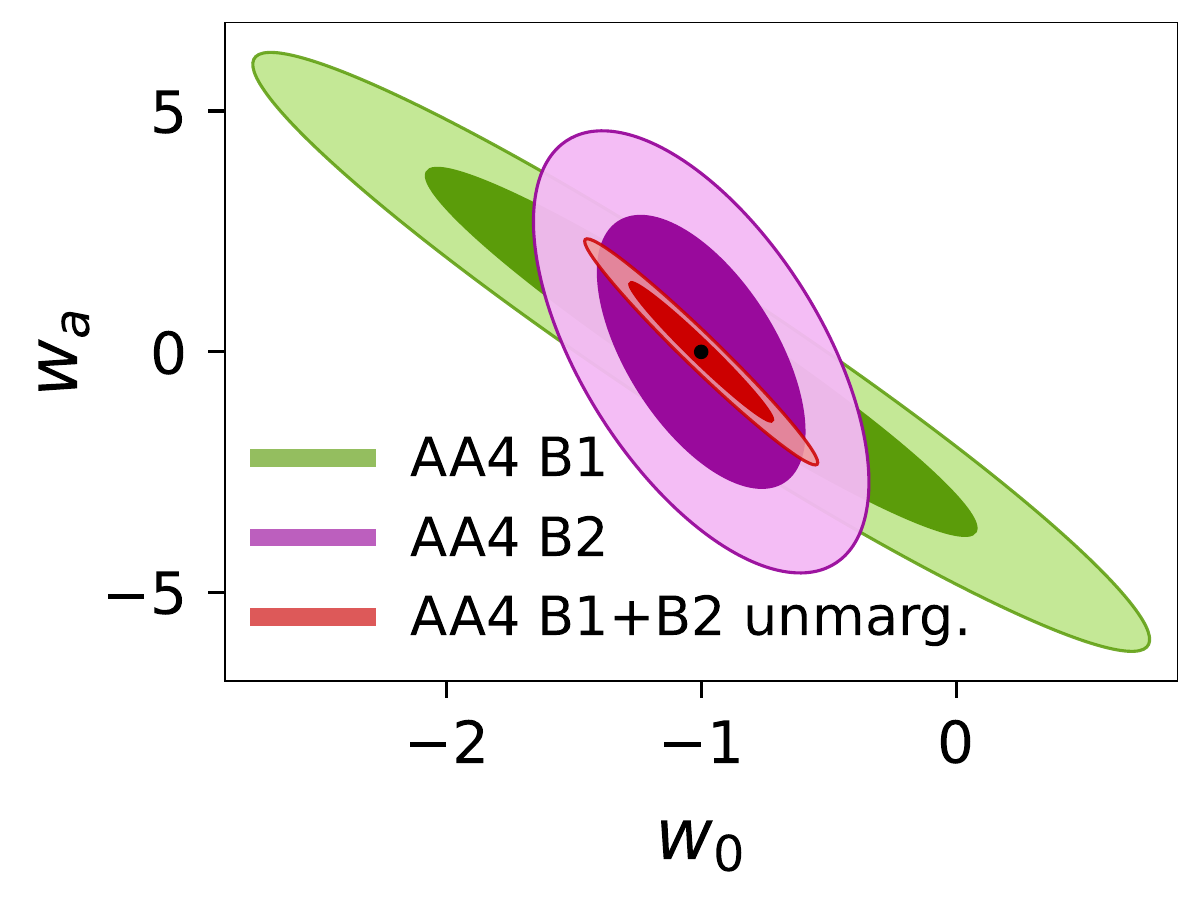}
    \caption{(Left): Forecast fractional error on $H(z)$ parameters with AA4 survey parameters and galaxy bias and number density values obtained from the S$^3$-SAX simulation. (Right): Forecast 1- and 2-$\sigma$ contours on the dark energy equation of state parameters $w_0,w_a$ for the same survey parameter/number density values.}
    \label{fig:waw0_comparison}
    \label{fig:hz_comparison}
  
\end{figure*}

Using the values for the bias $b$ and the number densities $dN/dz$ detailed above from S$^3$-SAX simulation and AA4 array configuration, we performed a forecast on the dark energy equation of state parameters $w_0, w_a$ shown in Fig.~\ref{fig:waw0_comparison} using the method and settings of \cite{Casas:2022vik}, where we adopt the conservative settings and nonlinear modeling of \cite{Euclid:2025tpw} for the H{\sc\,i} full shape power spectrum probe. We also left free the matter fluctuation parameter $\sigma_8$ and the matter density parameter $\Omega_m$, which we marginalise over. The rest of the cosmological parameters are kept fixed to their fiducial values however, which is highly optimistic in one sense. The purpose of excluding them here is to provide a rudimentary model of a joint analysis with other SKA or current CMB probes however -- since these parameters will be strongly constrained by the other surveys, whether they are fixed or not should not matter much in principle. (In reality, the correlations with other parameters should also be accounted for.)

Fig.~\ref{fig:waw0_comparison} shows $w_0, w_a$ confidence regions from Band 1 and Band 2 and their combination, suggesting that an SKA-Mid H{\sc\,i} galaxy survey will be able to reach 20\% constraints on $w_0$ and 40\% on $w_a$ with the support of other surveys.

We also performed a more model-agnostic forecast, keeping the same prescription, but considering a piecewise model for the expansion rate $H(z_i)$ in each redshift bin $i$ and similar for ${f\sigma_8}(z_i)$ upon which we marginalize at the end. Fig.~\ref{fig:hz_comparison} shows that, using the full shape H{\sc\,i} power spectrum (while fixing the remaining cosmological parameters), we are able to achieve strong (percent-level) constraints on $H(z)$, which are close to the ones obtained with only BAO + RSD as expected since the main constraints from the full shape comes from the BAO and RSD signature. This is also due to the fact that the BAO + RSD constraints are shown for Band 1 survey and Band 2 separately, rather than the combination of Band 1 and 2. Since $H(z_i)$ enters the comoving distance, it should correlate through more bins with this approach, hence becoming better constrained, despite introducing more degrees of freedom, especially at redshifts at the middle of our binning scheme range.

\subsection{Forecasts from the TO scale}

As a TO scale measurement requires to probe very large volumes, we consider here all H{\sc\,i} galaxies in a single redshift bin. We largely follow the \texttt{turnover\_examples} notebook in the DESI likelihood framework \texttt{desilike}\footnote{\url{https://github.com/cosmodesi/desilike/blob/main/nb/turnover_examples.ipynb}}. We start with computing the effective redshift $z_\mathrm{eff} = 0.23$ and H{\sc\,i} bias $b_\mathrm{HI,eff} = 0.92$ from the TNG100 simulations. We compute the matter power spectrum $P_m(k, z)$ for a standard flat $\Lambda$CDM model defined by the DESI fiducial cosmology, sourced via the \texttt{cosmoprimo}\footnote{\url{https://github.com/cosmodesi/cosmoprimo}} library. The growth rate $f(z)$ was calculated self-consistently from the same library to compute the angle-averaged H\,\textsc{i} power spectrum integrating over Eq.~\ref{eq:Pkzmu}. The total comoving volume $V$ for the survey was determined by integrating the volume element over the redshift extent and the fractional sky coverage ($f_{\rm sky}=\mathbf{0.125}$). This volume was used to calculate the comoving number density, $\bar{n}_{\rm HI} = \frac{{N}_{\rm gal}}{V}$, and consequently the shot noise, $P_{\rm shot} = 1/\bar{n}_{\rm HI}$. A diagonal Gaussian covariance matrix, $C_{ij}$, was constructed assuming uncorrelated $k$-bins. The variance is given by
\begin{equation}
C_{ij}(k_i) = \frac{2\, (2\pi)^3}{V N_{\rm modes}(k_i)} \left[P(k_i)\right]^2\delta_{ij}^\mathrm{K}\;,
\end{equation}
where $N_{\rm modes}(k_i)$ is the number of modes in the $i$-th $k$-shell and $\delta_{ij}^\mathrm{K}$ is the Kronecker delta symbol. One thousand Gaussian mock realisations were generated using the calculated mean multipoles and the covariance matrix to provide a robust estimate of the statistical uncertainty on the observables.

We isolate the TO scale using the parametrisation first proposed by \citet{2005MNRAS.363.1329B} and further refined by \citet{2013MNRAS.429.1902P,2023MNRAS.524.2463B} and \citet{2025PhRvD.112f3553B}, where the power spectrum left (right) of the peak is proportional to $P_\mathrm{TO, fid}^{1 - m x^2}$ ($P_\mathrm{TO, fid}^{1 - n x^2}$), where $k_\mathrm{TO,fid}$ and $P_\mathrm{TO, fid}$ are the turnover position and the power spectrum amplitude in the fiducial cosmology, $m$ and $n$ are slope parameters to be fitted, and
\begin{equation}
x = \frac{\ln(k\;\mathrm{Mpc}/h)}{\ln\left(\alpha_\mathrm{TO}k_\mathrm{TO,fid}\;\mathrm{Mpc}/h\right)} - 1.
\end{equation}
The parameter $\alpha_\mathrm{TO}$ shifts the turnover position with respect to the fiducial template and is also to be fitted. It is related to the quantities we want to measure as
\begin{equation}
\alpha_\mathrm{TO}^3 = \frac{{\Delta\theta_\mathrm{TO,fid}^2\Delta z_\mathrm{TO,fid}}}{{\Delta\theta_\mathrm{TO}^2\Delta z_\mathrm{TO}}},
\end{equation}
where the subscript `fid' again refers to the usual quantities evaluated in the fiducial cosmology. In summary, we fit four parameters: the peak position $\alpha_\mathrm{TO}$, two slopes $m$ and $n$, as well as the bias $b_\mathrm{HI,eff}$ as a proxy for the amplitude. 

We find $\sigma_{\alpha_\mathrm{TO}}/\alpha_\mathrm{TO} = 0.14$ for a single redshift bin. While this is higher than the forecast fractional errors on $H(z_\mathrm{eff})$ and $d_\mathrm{A}(z_\mathrm{eff})$ (cf. Fig. \ref{fig:hz_da}), the precision falls between the levels achieved by the eBOSS quasar measurement \citep{2023MNRAS.524.2463B} and the Gaia-unWISE quasar catalogue \citep{2025OJAp....8E..42A}. This result provides valuable low-redshift information complementary to higher-redshift probes. In particular, combining this measurement with existing high-redshift results will enable us to break the degeneracy between the particle horizon at matter-radiation equality and $\Omega_{m}$, as outlined in \citet{2025JCAP...10..071L}.

\section{Conclusions}

Optical galaxy redshift surveys have proven to be a powerful tool for observational cosmology, providing distance measures from the baryon acoustic oscillation and matter-radiation equality scales, and measurements of the cosmic growth rate from redshift-space distortions. Combined with other probes, such as cosmic microwave background power spectra, weak gravitational lensing surveys, and Type Ia supernovae, they have helped constrain the parameters of the standard cosmological model with high precision. Inconsistencies or `tensions' are now beginning to appear however, with different combinations of datasets preferring different and sometimes mutually-incompatible parameter values. This is most evident in measurements of the Hubble parameter, $H_{0}$, although estimates of other parameters such as $\Omega_{m}$ and $\sigma_8$ are arguably also in tension. Recently, the DESI survey reported a significant detection of time-varying dark energy from measurements of galaxy clustering at a range of redshifts, which could signify a deviation from the standard $\Lambda$CDM paradigm entirely.

A very high standard of evidence is required to conclude that the root causes of these tensions are physical, as it signifies a serious break in the prevailing cosmological paradigm. Other surveys and observational methods are needed to verify the tensions, and show that they are not the simple result of complicated systematic effects in the data or analysis methods. H{\sc\,i} galaxy surveys offer such a complementary data set. This can only be done by a wide area survey with the SKAO (mainly in Band 2). 

With such an H{\sc\,i} galaxy survey one can look at the statistics of the number density fluctuations to extract important cosmological information. With both the BAO and turn-over scales we will probe the expansion rate of the universe being able to attain percent level constrains. The RSDs are sensitive to the peculiar velocities of galaxies in large gravitational potential and, therefore, one will constrain the linear growth rate of structures in the universe. This in turn offers valuable information both on the theory of gravity. Both RSDs and the BAO scales will give complementary data points to probe the Dark Energy equation of state. 

Our forecasts are sensitive to the exact number of sources observed and how they trace the dark matter field. Here we have summarised the simulations available as well as their predictions for the size of the H{\sc\,i} galaxies catalogue. But the future catalogue provided by SKAO observations will constrain the number density and bias of H{\sc\,i} galaxies providing another complementary way to understand how gas has populated galaxies throughout the history of the Universe, and these the dark matter halos.

\section*{Author List Ordering}
The lead author coordinated the preparation of the chapter, performed one of the Fisher forecasts, and wrote Section 3 and most of Section 4. All other authors contributed to writing different sections, performing simulations, and/or providing guidance and useful comments, and they are listed in order of contribution.

\section*{Acknowledgments}
AN and PB acknowledge support from the European Research Council (ERC) under the European Union's Horizon 2020 research and innovation programme (Grant agreement No. 948764).
JM is supported in part by grant CRSII5\_193826 from the Swiss National Science Foundation, and by SERI as part of the SKACH consortium. BB-K acknowledges support from INAF for the project `Paving the way to radio cosmology in the SKA Observatory era: synergies between SKA pathfinders/precursors and the new generation of optical/near-infrared cosmological surveys', CUP C54I1900105 0001. LF acknowledges support from the South African National Research Foundation (NRF). ZS acknowledges support from the research projects PID2021-123012NB-C43, PID2024-159420NB-C43, the Proyecto de Investigación SAFE25003 from the Consejo Superior de Investigaciones Científicas (CSIC), and the Spanish Research Agency (Agencia Estatal de Investigaci\'on) through the Grant IFT Centro de Excelencia Severo Ochoa No CEX2020-001007-S, funded by MCIN/AEI/10.13039/501100011033. JF acknowledges support of Funda\c{c}\~{a}o para a Ci\^{e}ncia e a Tecnologia (FCT) through the Investigador FCT Contract no.\ 2020.02633.CEECIND/CP1631/CT0002, the FCT exploratory project 10.54499/2023.15069.PEX, and the research grant UID/04434/2025.

\bibliographystyle{abbrvnat-maxbibnames4}
\bibliography{chapter}

@techreport{Braun2024SKA1SciencePerformance,
  author      = {Braun, R. and et al.},
  title       = {Anticipated {SKA1} Science Performance},
  institution = {SKA Observatory},
  number      = {SKAO-TEL-0000818},
  type        = {DTE},
  version     = {Revision 002},
  year        = {2024},
  month       = oct,
  date        = {2024-10-08},
  url         = {https://www.skao.int/sites/default/files/documents/SKAO-TEL-0000818-V2_SKA1_Science_Performance.pdf},
  note        = {Classification: UNRESTRICTED},
}

@article{Pan_2020,
   title={Multiwavelength consensus of large-scale linear bias},
   volume={493},
   ISSN={1365-2966},
   url={http://dx.doi.org/10.1093/mnras/staa222},
   DOI={10.1093/mnras/staa222},
   number={1},
   journal={Monthly Notices of the Royal Astronomical Society},
   publisher={Oxford University Press (OUP)},
   author={Pan, Hengxing and Obreschkow, Danail and Howlett, Cullan and Lagos, Claudia del P and Elahi, Pascal J and Baugh, Carlton and Gonzalez-Perez, Violeta},
   year={2020},
   month=jan, pages={747–764} }

@article{Casas:2022vik,
    author = "Casas, Santiago and Carucci, Isabella P. and Pettorino, Valeria and Camera, Stefano and Martinelli, Matteo",
    title = "{Constraining gravity with synergies between radio and optical cosmological surveys}",
    eprint = "2210.05705",
    archivePrefix = "arXiv",
    primaryClass = "astro-ph.CO",
    doi = "10.1016/j.dark.2022.101151",
    journal = "Phys. Dark Univ.",
    volume = "39",
    pages = "101151",
    year = "2023"
}

@article{Euclid:2025tpw,
    author = "Albuquerque, I. S. and others",
    collaboration = "Euclid",
    title = "{Euclid preparation. Constraining parameterised models of modifications of gravity with the spectroscopic and photometric primary probes}",
    eprint = "2506.03008",
    archivePrefix = "arXiv",
    primaryClass = "astro-ph.CO",
    month = "6",
    year = "2025"
}

@ARTICLE{tng2018MNRAS.475..648P,
       author = {{Pillepich}, Annalisa and {Nelson}, Dylan and {Hernquist}, Lars and {Springel}, Volker and {Pakmor}, R{\"u}diger and {Torrey}, Paul and {Weinberger}, Rainer and {Genel}, Shy and {Naiman}, Jill P. and {Marinacci}, Federico and {Vogelsberger}, Mark},
        title = "{First results from the IllustrisTNG simulations: the stellar mass content of groups and clusters of galaxies}",
      journal = {\mnras},
         year = 2018,
        month = mar,
       volume = {475},
       number = {1},
        pages = {648-675},
          doi = {10.1093/mnras/stx3112},
archivePrefix = {arXiv},
       eprint = {1707.03406},
 primaryClass = {astro-ph.GA},
       adsurl = {https://ui.adsabs.harvard.edu/abs/2018MNRAS.475..648P}
}

@ARTICLE{tng2018MNRAS.475..624N,
       author = {{Nelson}, Dylan and {Pillepich}, Annalisa and {Springel}, Volker and {Weinberger}, Rainer and {Hernquist}, Lars and {Pakmor}, R{\"u}diger and {Genel}, Shy and {Torrey}, Paul and {Vogelsberger}, Mark and {Kauffmann}, Guinevere and {Marinacci}, Federico and {Naiman}, Jill},
        title = "{First results from the IllustrisTNG simulations: the galaxy colour bimodality}",
      journal = {\mnras},
         year = 2018,
        month = mar,
       volume = {475},
       number = {1},
        pages = {624-647},
          doi = {10.1093/mnras/stx3040},
archivePrefix = {arXiv},
       eprint = {1707.03395},
 primaryClass = {astro-ph.GA},
       adsurl = {https://ui.adsabs.harvard.edu/abs/2018MNRAS.475..624N}
}

@ARTICLE{tng2018MNRAS.475..676S,
       author = {{Springel}, Volker and {Pakmor}, R{\"u}diger and {Pillepich}, Annalisa and {Weinberger}, Rainer and {Nelson}, Dylan and {Hernquist}, Lars and {Vogelsberger}, Mark and {Genel}, Shy and {Torrey}, Paul and {Marinacci}, Federico and {Naiman}, Jill},
        title = "{First results from the IllustrisTNG simulations: matter and galaxy clustering}",
      journal = {\mnras},
         year = 2018,
        month = mar,
       volume = {475},
       number = {1},
        pages = {676-698},
          doi = {10.1093/mnras/stx3304},
archivePrefix = {arXiv},
       eprint = {1707.03397},
 primaryClass = {astro-ph.GA},
       adsurl = {https://ui.adsabs.harvard.edu/abs/2018MNRAS.475..676S}
}

@ARTICLE{tng2018MNRAS.480.5113M,
       author = {{Marinacci}, Federico and {Vogelsberger}, Mark and {Pakmor}, R{\"u}diger and {Torrey}, Paul and {Springel}, Volker and {Hernquist}, Lars and {Nelson}, Dylan and {Weinberger}, Rainer and {Pillepich}, Annalisa and {Naiman}, Jill and {Genel}, Shy},
        title = "{First results from the IllustrisTNG simulations: radio haloes and magnetic fields}",
      journal = {\mnras},
         year = 2018,
        month = nov,
       volume = {480},
       number = {4},
        pages = {5113-5139},
          doi = {10.1093/mnras/sty2206},
archivePrefix = {arXiv},
       eprint = {1707.03396},
 primaryClass = {astro-ph.CO},
       adsurl = {https://ui.adsabs.harvard.edu/abs/2018MNRAS.480.5113M}
}

@ARTICLE{tng2018MNRAS.477.1206N,
       author = {{Naiman}, Jill P. and {Pillepich}, Annalisa and {Springel}, Volker and {Ramirez-Ruiz}, Enrico and {Torrey}, Paul and {Vogelsberger}, Mark and {Pakmor}, R{\"u}diger and {Nelson}, Dylan and {Marinacci}, Federico and {Hernquist}, Lars and {Weinberger}, Rainer and {Genel}, Shy},
        title = "{First results from the IllustrisTNG simulations: a tale of two elements - chemical evolution of magnesium and europium}",
      journal = {\mnras},
         year = 2018,
        month = jun,
       volume = {477},
       number = {1},
        pages = {1206-1224},
          doi = {10.1093/mnras/sty618},
archivePrefix = {arXiv},
       eprint = {1707.03401},
 primaryClass = {astro-ph.GA},
       adsurl = {https://ui.adsabs.harvard.edu/abs/2018MNRAS.477.1206N}
}

@ARTICLE{tng2019ComAC...6....2N,
       author = {{Nelson}, Dylan and {Springel}, Volker and {Pillepich}, Annalisa and {Rodriguez-Gomez}, Vicente and {Torrey}, Paul and {Genel}, Shy and {Vogelsberger}, Mark and {Pakmor}, Ruediger and {Marinacci}, Federico and {Weinberger}, Rainer and {Kelley}, Luke and {Lovell}, Mark and {Diemer}, Benedikt and {Hernquist}, Lars},
        title = "{The IllustrisTNG simulations: public data release}",
      journal = {Computational Astrophysics and Cosmology},
         year = 2019,
        month = may,
       volume = {6},
       number = {1},
          eid = {2},
        pages = {2},
          doi = {10.1186/s40668-019-0028-x},
archivePrefix = {arXiv},
       eprint = {1812.05609},
 primaryClass = {astro-ph.GA},
       adsurl = {https://ui.adsabs.harvard.edu/abs/2019ComAC...6....2N}
}

@article{Diemer_2018,
doi = {10.3847/1538-4365/aae387},
url = {https://doi.org/10.3847/1538-4365/aae387},
year = {2018},
month = {oct},
publisher = {The American Astronomical Society},
volume = {238},
number = {2},
pages = {33},
author = {Diemer, Benedikt and Stevens, Adam R. H. and Forbes, John C. and Marinacci, Federico and Hernquist, Lars and Lagos, Claudia del P. and Sternberg, Amiel and Pillepich, Annalisa and Nelson, Dylan and Popping, Gergö and Villaescusa-Navarro, Francisco and Torrey, Paul and Vogelsberger, Mark},
title = {Modeling the Atomic-to-molecular Transition in Cosmological Simulations of Galaxy Formation},
journal = {The Astrophysical Journal Supplement Series}
}

@article{yahya10.1093/mnras/stv695,
    author = {Yahya, S. and Bull, P. and Santos, M. G. and Silva, M. and Maartens, R. and Okouma, P. and Bassett, B.},
    title = {Cosmological performance of SKA H I galaxy surveys},
    journal = {Monthly Notices of the Royal Astronomical Society},
    volume = {450},
    number = {3},
    pages = {2251-2260},
    year = {2015},
    month = {05},
    issn = {0035-8711},
    doi = {10.1093/mnras/stv695},
    url = {https://doi.org/10.1093/mnras/stv695},
    eprint = {https://academic.oup.com/mnras/article-pdf/450/3/2251/18504481/stv695.pdf},
}

@article{Spinelli:2021emp,
    author = "Spinelli, Marta and Carucci, Isabella P. and Cunnington, Steven and Harper, Stuart E. and Irfan, Melis O. and Fonseca, Jos{\'e} and Pourtsidou, Alkistis and Wolz, Laura",
    title = "{SKAO H{\,}i intensity mapping: blind foreground subtraction challenge}",
    eprint = "2107.10814",
    archivePrefix = "arXiv",
    primaryClass = "astro-ph.CO",
    doi = "10.1093/mnras/stab3064",
    journal = "Mon. Not. Roy. Astron. Soc.",
    volume = "509",
    number = "2",
    pages = "2048--2074",
    year = "2021"
}

@ARTICLE{2005Natur.435..629S,
       author = {{Springel}, Volker and {White}, Simon D.~M. and {Jenkins}, Adrian and {Frenk}, Carlos S. and {Yoshida}, Naoki and {Gao}, Liang and {Navarro}, Julio and {Thacker}, Robert and {Croton}, Darren and {Helly}, John and {Peacock}, John A. and {Cole}, Shaun and {Thomas}, Peter and {Couchman}, Hugh and {Evrard}, August and {Colberg}, J{\"o}rg and {Pearce}, Frazer},
        title = "{Simulations of the formation, evolution and clustering of galaxies and quasars}",
      journal = {\nat},
         year = 2005,
        month = jun,
       volume = {435},
       number = {7042},
        pages = {629-636},
          doi = {10.1038/nature03597},
archivePrefix = {arXiv},
       eprint = {astro-ph/0504097},
 primaryClass = {astro-ph},
       adsurl = {https://ui.adsabs.harvard.edu/abs/2005Natur.435..629S}
}

@ARTICLE{2007ApJ...664..675E,
       author = {{Eisenstein}, Daniel J. and {Seo}, Hee-Jong and {Sirko}, Edwin and {Spergel}, David N.},
        title = "{Improving Cosmological Distance Measurements by Reconstruction of the Baryon Acoustic Peak}",
      journal = {\apj},
         year = 2007,
        month = aug,
       volume = {664},
       number = {2},
        pages = {675-679},
          doi = {10.1086/518712},
archivePrefix = {arXiv},
       eprint = {astro-ph/0604362},
 primaryClass = {astro-ph},
       adsurl = {https://ui.adsabs.harvard.edu/abs/2007ApJ...664..675E}
}

@incollection{Ronconi01.2026.SKA, 
author = {Tommaso Ronconi and author2 and author3 and author4 and author5},
title = {},
year = {2026},
publisher = {},
note = {arXiv search: Report number AASKAII/Ronconi01},
booktitle = {Advancing Astrophysics with the SKA -- II (AASKAII)}}

@ARTICLE{2020PASA...37....7S,
       author = {{Square Kilometre Array Cosmology Science Working Group} and {Bacon}, David J. and {Battye}, Richard A. and {Bull}, Philip and {Camera}, Stefano and {Ferreira}, Pedro G. and {Harrison}, Ian and {Parkinson}, David and {Pourtsidou}, Alkistis and {Santos}, M{\'a}rio G. and {Wolz}, Laura and {Abdalla}, Filipe and {Akrami}, Yashar and {Alonso}, David and {Andrianomena}, Sambatra and {Ballardini}, Mario and {Bernal}, Jos{\'e} Luis and {Bertacca}, Daniele and {Bengaly}, Carlos A.~P. and {Bonaldi}, Anna and {Bonvin}, Camille and {Brown}, Michael L. and {Chapman}, Emma and {Chen}, Song and {Chen}, Xuelei and {Cunnington}, Steven and {Davis}, Tamara M. and {Dickinson}, Clive and {Fonseca}, Jos{\'e} and {Grainge}, Keith and {Harper}, Stuart and {Jarvis}, Matt J. and {Maartens}, Roy and {Maddox}, Natasha and {Padmanabhan}, Hamsa and {Pritchard}, Jonathan R. and {Raccanelli}, Alvise and {Rivi}, Marzia and {Roychowdhury}, Sambit and {Sahl{\'e}n}, Martin and {Schwarz}, Dominik J. and {Siewert}, Thilo M. and {Viel}, Matteo and {Villaescusa-Navarro}, Francisco and {Xu}, Yidong and {Yamauchi}, Daisuke and {Zuntz}, Joe},
        title = "{Cosmology with Phase 1 of the Square Kilometre Array Red Book 2018: Technical specifications and performance forecasts}",
      journal = {\pasa},
         year = 2020,
        month = mar,
       volume = {37},
          eid = {e007},
        pages = {e007},
          doi = {10.1017/pasa.2019.51},
archivePrefix = {arXiv},
       eprint = {1811.02743},
 primaryClass = {astro-ph.CO},
       adsurl = {https://ui.adsabs.harvard.edu/abs/2020PASA...37....7S}
}

@ARTICLE{2012arXiv1212.3497E,
       author = {{Ekers}, Ron},
        title = "{The History of the Square Kilometre Array (SKA) - Born Global}",
      journal = {arXiv e-prints},
         year = 2012,
        month = dec,
          eid = {arXiv:1212.3497},
        pages = {arXiv:1212.3497},
          doi = {10.48550/arXiv.1212.3497},
archivePrefix = {arXiv},
       eprint = {1212.3497},
 primaryClass = {astro-ph.IM},
       adsurl = {https://ui.adsabs.harvard.edu/abs/2012arXiv1212.3497E}
}

@ARTICLE{2017MNRAS.470..340P,
       author = {{Padmanabhan}, Hamsa and {Kulkarni}, Girish},
        title = "{Constraints on the evolution of the relationship between H I mass and halo mass in the last 12 Gyr}",
      journal = {\mnras},
         year = 2017,
        month = sep,
       volume = {470},
       number = {1},
        pages = {340-349},
          doi = {10.1093/mnras/stx1178},
archivePrefix = {arXiv},
       eprint = {1608.00007},
 primaryClass = {astro-ph.GA},
       adsurl = {https://ui.adsabs.harvard.edu/abs/2017MNRAS.470..340P}
}

@ARTICLE{2017MNRAS.471.1788C,
       author = {{Castorina}, Emanuele and {Villaescusa-Navarro}, Francisco},
        title = "{On the spatial distribution of neutral hydrogen in the Universe: bias and shot-noise of the H I power spectrum}",
      journal = {\mnras},
         year = 2017,
        month = oct,
       volume = {471},
       number = {2},
        pages = {1788-1796},
          doi = {10.1093/mnras/stx1599},
archivePrefix = {arXiv},
       eprint = {1609.05157},
 primaryClass = {astro-ph.CO},
       adsurl = {https://ui.adsabs.harvard.edu/abs/2017MNRAS.471.1788C}
}

@ARTICLE{2025RSPTA.38340022E,
       author = {{Efstathiou}, George},
        title = "{Challenges to the {\ensuremath{\Lambda}} CDM cosmology}",
      journal = {Philosophical Transactions of the Royal Society of London Series A},
         year = 2025,
        month = feb,
       volume = {383},
       number = {2290},
          eid = {20240022},
        pages = {20240022},
          doi = {10.1098/rsta.2024.0022},
archivePrefix = {arXiv},
       eprint = {2406.12106},
 primaryClass = {astro-ph.CO},
       adsurl = {https://ui.adsabs.harvard.edu/abs/2025RSPTA.38340022E}
}

@ARTICLE{2021CQGra..38o3001D,
       author = {{Di Valentino}, Eleonora and {Mena}, Olga and {Pan}, Supriya and {Visinelli}, Luca and {Yang}, Weiqiang and {Melchiorri}, Alessandro and {Mota}, David F. and {Riess}, Adam G. and {Silk}, Joseph},
        title = "{In the realm of the Hubble tension-a review of solutions}",
      journal = {Classical and Quantum Gravity},
         year = 2021,
        month = jul,
       volume = {38},
       number = {15},
          eid = {153001},
        pages = {153001},
          doi = {10.1088/1361-6382/ac086d},
archivePrefix = {arXiv},
       eprint = {2103.01183},
 primaryClass = {astro-ph.CO},
       adsurl = {https://ui.adsabs.harvard.edu/abs/2021CQGra..38o3001D}
}

@ARTICLE{2016A&A...593A..39P,
       author = {{Papastergis}, E. and {Adams}, E.~A.~K. and {van der Hulst}, J.~M.},
        title = "{An accurate measurement of the baryonic Tully-Fisher relation with heavily gas-dominated ALFALFA galaxies}",
      journal = {\aap},
         year = 2016,
        month = sep,
       volume = {593},
          eid = {A39},
        pages = {A39},
          doi = {10.1051/0004-6361/201628410},
archivePrefix = {arXiv},
       eprint = {1602.09087},
 primaryClass = {astro-ph.GA},
       adsurl = {https://ui.adsabs.harvard.edu/abs/2016A&A...593A..39P}
}

@ARTICLE{2008MNRAS.391.1712M,
       author = {{Meyer}, M.~J. and {Zwaan}, M.~A. and {Webster}, R.~L. and {Schneider}, S. and {Staveley-Smith}, L.},
        title = "{Tully-Fisher relations from an HI-selected sample}",
      journal = {\mnras},
         year = 2008,
        month = dec,
       volume = {391},
       number = {4},
        pages = {1712-1728},
          doi = {10.1111/j.1365-2966.2008.13424.x},
       adsurl = {https://ui.adsabs.harvard.edu/abs/2008MNRAS.391.1712M}
}

@ARTICLE{2024MNRAS.533.1550B,
       author = {{Boubel}, Paula and {Colless}, Matthew and {Said}, Khaled and {Staveley-Smith}, Lister},
        title = "{An improved Tully-Fisher estimate of H$_{0}$}",
      journal = {\mnras},
         year = 2024,
        month = sep,
       volume = {533},
       number = {2},
        pages = {1550-1559},
          doi = {10.1093/mnras/stae1925},
archivePrefix = {arXiv},
       eprint = {2408.03660},
 primaryClass = {astro-ph.CO},
       adsurl = {https://ui.adsabs.harvard.edu/abs/2024MNRAS.533.1550B}
}

@ARTICLE{2023ApJ...950...87B,
       author = {{Ball}, Catie J. and {Haynes}, Martha P. and {Jones}, Michael G. and {Peng}, Bo and {Durbala}, Adriana and {Koopmann}, Rebecca A. and {Ribaudo}, Joseph and {O'Donoghue}, Aileen A.},
        title = "{A Generalist, Automated ALFALFA Baryonic Tully-Fisher Relation}",
      journal = {\apj},
         year = 2023,
        month = jun,
       volume = {950},
       number = {2},
          eid = {87},
        pages = {87},
          doi = {10.3847/1538-4357/accb53},
archivePrefix = {arXiv},
       eprint = {2212.08728},
 primaryClass = {astro-ph.GA},
       adsurl = {https://ui.adsabs.harvard.edu/abs/2023ApJ...950...87B}
}

@ARTICLE{2025PhRvD.112h3515A,
       author = {{Abdul Karim}, M. and {Aguilar}, J. and {Ahlen}, S. and {Alam}, S. and {Allen}, L. and {Prieto}, C. Allende and {Alves}, O. and {Anand}, A. and {Andrade}, U. and {Armengaud}, E. and {Aviles}, A. and {Bailey}, S. and {Baltay}, C. and {Bansal}, P. and {Bault}, A. and {Behera}, J. and {BenZvi}, S. and {Bianchi}, D. and {Blake}, C. and {Brieden}, S. and {Brodzeller}, A. and {Brooks}, D. and {Buckley-Geer}, E. and {Burtin}, E. and {Calderon}, R. and {Canning}, R. and {Rosell}, A. Carnero and {Carrilho}, P. and {Casas}, L. and {Castander}, F.~J. and {Charles}, M. and {Chaussidon}, E. and {Chaves-Montero}, J. and {Chebat}, D. and {Chen}, X. and {Claybaugh}, T. and {Cole}, S. and {Cooper}, A.~P. and {Cuceu}, A. and {Dawson}, K.~S. and {de la Macorra}, A. and {de Mattia}, A. and {Deiosso}, N. and {Della Costa}, J. and {Demina}, R. and {Dey}, A. and {Dey}, B. and {Ding}, Z. and {Doel}, P. and {Edelstein}, J. and {Eisenstein}, D.~J. and {Elbers}, W. and {Fagrelius}, P. and {Fanning}, K. and {Fern{\'a}ndez-Garc{\'\i}a}, E. and {Ferraro}, S. and {Font-Ribera}, A. and {Forero-Romero}, J.~E. and {Frenk}, C.~S. and {Garcia-Quintero}, C. and {Garrison}, L.~H. and {Gazta{\~n}aga}, E. and {Gil-Mar{\'\i}n}, H. and {Gontcho A Gontcho}, S. and {Gonzalez}, D. and {Gonzalez-Morales}, A.~X. and {Gordon}, C. and {Green}, D. and {Gutierrez}, G. and {Guy}, J. and {Hadzhiyska}, B. and {Hahn}, C. and {He}, S. and {Herbold}, M. and {Herrera-Alcantar}, H.~K. and {Ho}, M.-F. and {Honscheid}, K. and {Howlett}, C. and {Huterer}, D. and {Ishak}, M. and {Juneau}, S. and {Kamble}, N.~V. and {Kara{\c{c}}ayl{\i}}, N.~G. and {Kehoe}, R. and {Kent}, S. and {Kim}, A.~G. and {Kirkby}, D. and {Kisner}, T. and {Koposov}, S.~E. and {Kremin}, A. and {Krolewski}, A. and {Lahav}, O. and {Lamman}, C. and {Landriau}, M. and {Lang}, D. and {Lasker}, J. and {Le Goff}, J.~M. and {Le Guillou}, L. and {Leauthaud}, A. and {Levi}, M.~E. and {Li}, Q. and {Li}, T.~S. and {Lodha}, K. and {Lokken}, M. and {Lozano-Rodr{\'\i}guez}, F. and {Magneville}, C. and {Manera}, M. and {Martini}, P. and {Matthewson}, W.~L. and {Meisner}, A. and {Mena-Fern{\'a}ndez}, J. and {Menegas}, A. and {Mergulh{\~a}o}, T. and {Miquel}, R. and {Moustakas}, J. and {Mu{\~n}oz-Guti{\'e}rrez}, A. and {Mu{\~n}oz-Santos}, D. and {Myers}, A.~D. and {Nadathur}, S. and {Naidoo}, K. and {Napolitano}, L. and {Newman}, J.~A. and {Niz}, G. and {Noriega}, H.~E. and {Paillas}, E. and {Palanque-Delabrouille}, N. and {Pan}, J. and {Peacock}, J.~A. and {Pellejero Ibanez}, M. and {Percival}, W.~J. and {P{\'e}rez-Fern{\'a}ndez}, A. and {P{\'e}rez-R{\`a}fols}, I. and {Pieri}, M.~M. and {Poppett}, C. and {Prada}, F. and {Rabinowitz}, D. and {Raichoor}, A. and {Ram{\'\i}rez-P{\'e}rez}, C. and {Rashkovetskyi}, M. and {Ravoux}, C. and {Rich}, J. and {Rocher}, A. and {Rockosi}, C. and {Rohlf}, J. and {Rom{\'a}n-Herrera}, J.~O. and {Ross}, A.~J. and {Rossi}, G. and {Ruggeri}, R. and {Ruhlmann-Kleider}, V. and {Samushia}, L. and {Sanchez}, E. and {Sanders}, N. and {Schlegel}, D. and {Schubnell}, M. and {Seo}, H. and {Shafieloo}, A. and {Sharples}, R. and {Silber}, J. and {Sinigaglia}, F. and {Sprayberry}, D. and {Tan}, T. and {Tarl{\'e}}, G. and {Taylor}, P. and {Turner}, W. and {Ure{\~n}a-L{\'o}pez}, L.~A. and {Vaisakh}, R. and {Valdes}, F. and {Valogiannis}, G. and {Vargas-Maga{\~n}a}, M. and {Verde}, L. and {Walther}, M. and {Weaver}, B.~A. and {Weinberg}, D.~H. and {White}, M. and {Wolfson}, M. and {Y{\`e}che}, C. and {Yu}, J. and {Zaborowski}, E.~A. and {Zarrouk}, P. and {Zhai}, Z. and {Zhang}, H. and {Zhao}, C. and {Zhao}, G.~B. and {Zhou}, R. and {Zou}, H. and {DESI Collaboration}},
        title = "{DESI DR2 results. II. Measurements of baryon acoustic oscillations and cosmological constraints}",
      journal = {\prd},
         year = 2025,
        month = oct,
       volume = {112},
       number = {8},
          eid = {083515},
        pages = {083515},
          doi = {10.1103/tr6y-kpc6},
archivePrefix = {arXiv},
       eprint = {2503.14738},
 primaryClass = {astro-ph.CO},
       adsurl = {https://ui.adsabs.harvard.edu/abs/2025PhRvD.112h3515A}
}

@ARTICLE{2021ApJ...919...16F,
       author = {{Freedman}, Wendy L.},
        title = "{Measurements of the Hubble Constant: Tensions in Perspective}",
      journal = {\apj},
         year = 2021,
        month = sep,
       volume = {919},
       number = {1},
          eid = {16},
        pages = {16},
          doi = {10.3847/1538-4357/ac0e95},
archivePrefix = {arXiv},
       eprint = {2106.15656},
 primaryClass = {astro-ph.CO},
       adsurl = {https://ui.adsabs.harvard.edu/abs/2021ApJ...919...16F}
}

@ARTICLE{2019MNRAS.486.5124O,
       author = {{Obuljen}, Andrej and {Alonso}, David and {Villaescusa-Navarro}, Francisco and {Yoon}, Ilsang and {Jones}, Michael},
        title = "{The H I content of dark matter haloes at z {\ensuremath{\approx}} 0 from ALFALFA}",
      journal = {\mnras},
         year = 2019,
        month = jul,
       volume = {486},
       number = {4},
        pages = {5124-5138},
          doi = {10.1093/mnras/stz1118},
archivePrefix = {arXiv},
       eprint = {1805.00934},
 primaryClass = {astro-ph.CO},
       adsurl = {https://ui.adsabs.harvard.edu/abs/2019MNRAS.486.5124O}
}

@ARTICLE{2016ApJ...817...26B,
       author = {{Bull}, Philip},
        title = "{Extending Cosmological Tests of General Relativity with the Square Kilometre Array}",
      journal = {\apj},
         year = 2016,
        month = jan,
       volume = {817},
       number = {1},
          eid = {26},
        pages = {26},
          doi = {10.3847/0004-637X/817/1/26},
archivePrefix = {arXiv},
       eprint = {1509.07562},
 primaryClass = {astro-ph.CO},
       adsurl = {https://ui.adsabs.harvard.edu/abs/2016ApJ...817...26B}
}

@article{10.1093/mnras/stae2090,
    author = {Chen, S -F and Howlett, C and White, M and McDonald, P and Ross, A J and Seo, H -J and Padmanabhan, N and Aguilar, J and Ahlen, S and Alam, S and Alves, O and Andrade, U and Blum, R and Brooks, D and Chen, X and Cole, S and Dawson, K and de la Macorra, A and Dey, A and Ding, Z and Doel, P and Ferraro, S and Font-Ribera, A and Forero-Sánchez, D and Forero-Romero, J E and Garcia-Quintero, C and Gaztañaga, E and Gontcho, S G A and Hanif, M M S and Honscheid, K and Kisner, T and Kremin, A and Lambert, A and Landriau, M and Levi, M E and Manera, M and Meisner, A and Mena-Fernández, J and Miquel, R and Munoz-Gutierrez, A and Paillas, E and Palanque-Delabrouille, N and Percival, W J and Pérez-Fernández, A and Prada, F and Rashkovetskyi, M and Rezaie, M and Rosado-Marin, A and Rossi, G and Ruggeri, R and Sanchez, E and Schlegel, D and Silber, J and Tarlé, G and Vargas-Magaña, M and Weaver, B A and Yu, J and Yuan, S and Zhou, R and Zhou, Z},
    title = {Baryon acoustic oscillation theory and modelling systematics for the DESI 2024 results},
    journal = {Monthly Notices of the Royal Astronomical Society},
    volume = {534},
    number = {1},
    pages = {544-574},
    year = {2024},
    month = {09},
    issn = {0035-8711},
    doi = {10.1093/mnras/stae2090},
    url = {https://doi.org/10.1093/mnras/stae2090},
    eprint = {https://academic.oup.com/mnras/article-pdf/534/1/544/59212495/stae2090.pdf},
}

@INPROCEEDINGS{santos2015aska.confE..21S,
       author = {{Santos}, M. and {Alonso}, D. and {Bull}, P. and {Silva}, M.~B. and {Yahya}, S.},
        title = "{HI galaxy simulations for the SKA: number counts and bias}",
    booktitle = {Advancing Astrophysics with the Square Kilometre Array (AASKA14)},
         year = 2015,
        month = apr,
          eid = {21},
        pages = {21},
          doi = {10.22323/1.215.0021},
archivePrefix = {arXiv},
       eprint = {1501.03990},
 primaryClass = {astro-ph.CO},
       adsurl = {https://ui.adsabs.harvard.edu/abs/2015aska.confE..21S}
}

@article{Bull_2015,
doi = {10.1088/0004-637X/803/1/21},
url = {https://doi.org/10.1088/0004-637X/803/1/21},
year = {2015},
month = {apr},
publisher = {The American Astronomical Society},
volume = {803},
number = {1},
pages = {21},
author = {Bull, Philip and Ferreira, Pedro G. and Patel, Prina and Santos, Mário G.},
title = {LATE-TIME COSMOLOGY WITH 21 cm INTENSITY MAPPING EXPERIMENTS},
journal = {The Astrophysical Journal}
}

@article{Meyer_Robotham_Obreschkow_Westmeier_Duffy_Staveley-Smith_2017, title={Tracing Hi Beyond the Local Universe},
volume={34},
DOI={10.1017/pasa.2017.31},
journal={Publications of the Astronomical Society of Australia},
author={Meyer, Martin and Robotham, Aaron and Obreschkow, Danail and Westmeier, Tobias and Duffy, Alan R. and Staveley-Smith, Lister},
year={2017},
pages={e052}}

@article{ refId0,
	author = {{Planck Collaboration} and {Ade, P. A. R.} and {Aghanim, N.} and {Arnaud, M.} and {Ashdown, M.} and {Aumont, J.} and {Baccigalupi, C.} and {Banday, A. J.} and {Barreiro, R. B.} and {Bartlett, J. G.} and {Bartolo, N.} and {Battaner, E.} and {Battye, R.} and {Benabed, K.} and {Benoît, A.} and {Benoit-Lévy, A.} and {Bernard, J.-P.} and {Bersanelli, M.} and {Bielewicz, P.} and {Bock, J. J.} and {Bonaldi, A.} and {Bonavera, L.} and {Bond, J. R.} and {Borrill, J.} and {Bouchet, F. R.} and {Boulanger, F.} and {Bucher, M.} and {Burigana, C.} and {Butler, R. C.} and {Calabrese, E.} and {Cardoso, J.-F.} and {Catalano, A.} and {Challinor, A.} and {Chamballu, A.} and {Chary, R.-R.} and {Chiang, H. C.} and {Chluba, J.} and {Christensen, P. R.} and {Church, S.} and {Clements, D. L.} and {Colombi, S.} and {Colombo, L. P. L.} and {Combet, C.} and {Coulais, A.} and {Crill, B. P.} and {Curto, A.} and {Cuttaia, F.} and {Danese, L.} and {Davies, R. D.} and {Davis, R. J.} and {de Bernardis, P.} and {de Rosa, A.} and {de Zotti, G.} and {Delabrouille, J.} and {Désert, F.-X.} and {Di Valentino, E.} and {Dickinson, C.} and {Diego, J. M.} and {Dolag, K.} and {Dole, H.} and {Donzelli, S.} and {Doré, O.} and {Douspis, M.} and {Ducout, A.} and {Dunkley, J.} and {Dupac, X.} and {Efstathiou, G.} and {Elsner, F.} and {Enßlin, T. A.} and {Eriksen, H. K.} and {Farhang, M.} and {Fergusson, J.} and {Finelli, F.} and {Forni, O.} and {Frailis, M.} and {Fraisse, A. A.} and {Franceschi, E.} and {Frejsel, A.} and {Galeotta, S.} and {Galli, S.} and {Ganga, K.} and {Gauthier, C.} and {Gerbino, M.} and {Ghosh, T.} and {Giard, M.} and {Giraud-Héraud, Y.} and {Giusarma, E.} and {Gjerløw, E.} and {González-Nuevo, J.} and {Górski, K. M.} and {Gratton, S.} and {Gregorio, A.} and {Gruppuso, A.} and {Gudmundsson, J. E.} and {Hamann, J.} and {Hansen, F. K.} and {Hanson, D.} and {Harrison, D. L.} and {Helou, G.} and {Henrot-Versillé, S.} and {Hernández-Monteagudo, C.} and {Herranz, D.} and {Hildebrandt, S. R.} and {Hivon, E.} and {Hobson, M.} and {Holmes, W. A.} and {Hornstrup, A.} and {Hovest, W.} and {Huang, Z.} and {Huffenberger, K. M.} and {Hurier, G.} and {Jaffe, A. H.} and {Jaffe, T. R.} and {Jones, W. C.} and {Juvela, M.} and {Keihänen, E.} and {Keskitalo, R.} and {Kisner, T. S.} and {Kneissl, R.} and {Knoche, J.} and {Knox, L.} and {Kunz, M.} and {Kurki-Suonio, H.} and {Lagache, G.} and {Lähteenmäki, A.} and {Lamarre, J.-M.} and {Lasenby, A.} and {Lattanzi, M.} and {Lawrence, C. R.} and {Leahy, J. P.} and {Leonardi, R.} and {Lesgourgues, J.} and {Levrier, F.} and {Lewis, A.} and {Liguori, M.} and {Lilje, P. B.} and {Linden-Vørnle, M.} and {López-Caniego, M.} and {Lubin, P. M.} and {Macías-Pérez, J. F.} and {Maggio, G.} and {Maino, D.} and {Mandolesi, N.} and {Mangilli, A.} and {Marchini, A.} and {Maris, M.} and {Martin, P. G.} and {Martinelli, M.} and {Martínez-González, E.} and {Masi, S.} and {Matarrese, S.} and {McGehee, P.} and {Meinhold, P. R.} and {Melchiorri, A.} and {Melin, J.-B.} and {Mendes, L.} and {Mennella, A.} and {Migliaccio, M.} and {Millea, M.} and {Mitra, S.} and {Miville-Deschênes, M.-A.} and {Moneti, A.} and {Montier, L.} and {Morgante, G.} and {Mortlock, D.} and {Moss, A.} and {Munshi, D.} and {Murphy, J. A.} and {Naselsky, P.} and {Nati, F.} and {Natoli, P.} and {Netterfield, C. B.} and {Nørgaard-Nielsen, H. U.} and {Noviello, F.} and {Novikov, D.} and {Novikov, I.} and {Oxborrow, C. A.} and {Paci, F.} and {Pagano, L.} and {Pajot, F.} and {Paladini, R.} and {Paoletti, D.} and {Partridge, B.} and {Pasian, F.} and {Patanchon, G.} and {Pearson, T. J.} and {Perdereau, O.} and {Perotto, L.} and {Perrotta, F.} and {Pettorino, V.} and {Piacentini, F.} and {Piat, M.} and {Pierpaoli, E.} and {Pietrobon, D.} and {Plaszczynski, S.} and {Pointecouteau, E.} and {Polenta, G.} and {Popa, L.} and {Pratt, G. W.} and {Prézeau, G.} and {Prunet, S.} and {Puget, J.-L.} and {Rachen, J. P.} and {Reach, W. T.} and {Rebolo, R.} and {Reinecke, M.} and {Remazeilles, M.} and {Renault, C.} and {Renzi, A.} and {Ristorcelli, I.} and {Rocha, G.} and {Rosset, C.} and {Rossetti, M.} and {Roudier, G.} and {Rouillé d’Orfeuil, B.} and {Rowan-Robinson, M.} and {Rubiño-Martín, J. A.} and {Rusholme, B.} and {Said, N.} and {Salvatelli, V.} and {Salvati, L.} and {Sandri, M.} and {Santos, D.} and {Savelainen, M.} and {Savini, G.} and {Scott, D.} and {Seiffert, M. D.} and {Serra, P.} and {Shellard, E. P. S.} and {Spencer, L. D.} and {Spinelli, M.} and {Stolyarov, V.} and {Stompor, R.} and {Sudiwala, R.} and {Sunyaev, R.} and {Sutton, D.} and {Suur-Uski, A.-S.} and {Sygnet, J.-F.} and {Tauber, J. A.} and {Terenzi, L.} and {Toffolatti, L.} and {Tomasi, M.} and {Tristram, M.} and {Trombetti, T.} and {Tucci, M.} and {Tuovinen, J.} and {Türler, M.} and {Umana, G.} and {Valenziano, L.} and {Valiviita, J.} and {Van Tent, F.} and {Vielva, P.} and {Villa, F.} and {Wade, L. A.} and {Wandelt, B. D.} and {Wehus, I. K.} and {White, M.} and {White, S. D. M.} and {Wilkinson, A.} and {Yvon, D.} and {Zacchei, A.} and {Zonca, A.}},
	title = {Planck 2015 results - XIII. Cosmological parameters},
	DOI= "10.1051/0004-6361/201525830",
	url= "https://doi.org/10.1051/0004-6361/201525830",
	journal = {\aap},
	year = 2016,
	volume = 594,
	pages = "A13",
}

@ARTICLE{1985ApJS...58...67T,
       author = {{Tully}, R.~B. and {Fouque}, P.},
        title = "{The extragalactic distance scale. I - Corrections to fundamental observables.}",
      journal = {\apjs},
         year = 1985,
        month = may,
       volume = {58},
        pages = {67-80},
          doi = {10.1086/191029},
       adsurl = {https://ui.adsabs.harvard.edu/abs/1985ApJS...58...67T}
}

@article{Obreschkow_2009b,
doi = {10.1088/0004-637X/703/2/1890},
url = {https://dx.doi.org/10.1088/0004-637X/703/2/1890},
year = {2009},
month = {sep},
publisher = {The American Astronomical Society},
volume = {703},
number = {2},
pages = {1890},
author = {Obreschkow, D. and Klöckner, H.-R. and Heywood, I. and Levrier, F. and Rawlings, S.},
title = {A VIRTUAL SKY WITH EXTRAGALACTIC H I AND CO LINES FOR THE SQUARE KILOMETRE ARRAY AND THE ATACAMA LARGE MILLIMETER/SUBMILLIMETER ARRAY*},
journal = {The Astrophysical Journal}
}

@article{duffy2012askap,
    author = {Duffy, Alan R. and Meyer, Martin J. and Staveley-Smith, Lister and Bernyk, Maksym and Croton, Darren J. and Koribalski, Bärbel S. and Gerstmann, Derek and Westerlund, Stefan},
    title = {Predictions for ASKAP neutral hydrogen surveys},
    journal = {Monthly Notices of the Royal Astronomical Society},
    volume = {426},
    number = {4},
    pages = {3385-3402},
    year = {2012},
    month = {11},
    issn = {0035-8711},
    doi = {10.1111/j.1365-2966.2012.21987.x},
    url = {https://doi.org/10.1111/j.1365-2966.2012.21987.x},
    eprint = {https://academic.oup.com/mnras/article-pdf/426/4/3385/3338965/426-4-3385.pdf},
}

@article{diemer2019tng,
    author = {Diemer, Benedikt and Stevens, Adam R H and Lagos, Claudia del P and Calette, A R and Tacchella, Sandro and Hernquist, Lars and Marinacci, Federico and Nelson, Dylan and Pillepich, Annalisa and Rodriguez-Gomez, Vicente and Villaescusa-Navarro, Francisco and Vogelsberger, Mark},
    title = {Atomic and molecular gas in IllustrisTNG galaxies at low redshift},
    journal = {Monthly Notices of the Royal Astronomical Society},
    volume = {487},
    number = {2},
    pages = {1529-1550},
    year = {2019},
    month = {05},
    issn = {0035-8711},
    doi = {10.1093/mnras/stz1323},
    url = {https://doi.org/10.1093/mnras/stz1323},
    eprint = {https://academic.oup.com/mnras/article-pdf/487/2/1529/28766047/stz1323.pdf},
}

@ARTICLE{2025PhRvD.112f3553B,
       author = {{Bahr-Kalus}, B. and {Parkinson}, D. and {Lodha}, K. and {Mueller}, E. and {Chaussidon}, E. and {de Mattia}, A. and {Forero-S{\'a}nchez}, D. and {Aguilar}, J. and {Ahlen}, S. and {Bianchi}, D. and {Brooks}, D. and {Claybaugh}, T. and {Cuceu}, A. and {de la Macorra}, A. and {Doel}, P. and {Font-Ribera}, A. and {Gazta{\~n}aga}, E. and {Gontcho}, S. Gontcho A. and {Gutierrez}, G. and {Honscheid}, K. and {Huterer}, D. and {Ishak}, M. and {Kehoe}, R. and {Kent}, S. and {Kirkby}, D. and {Kisner}, T. and {Kremin}, A. and {Lahav}, O. and {Landriau}, M. and {Le Guillou}, L. and {Magneville}, C. and {Manera}, M. and {Martini}, P. and {Meisner}, A. and {Miquel}, R. and {Moustakas}, J. and {Nadathur}, S. and {Palanque-Delabrouille}, N. and {Percival}, W.~J. and {Prada}, F. and {P{\'e}rez-R{\`a}fols}, I. and {Ross}, A.~J. and {Rossi}, G. and {Samushia}, L. and {Sanchez}, E. and {Schlegel}, D. and {Schubnell}, M. and {Seo}, H. and {Silber}, J. and {Sprayberry}, D. and {Tarl{\'e}}, G. and {Weaver}, B.~A. and {Zhou}, R. and {Zou}, H. and {DESI Collaboration}},
        title = "{Model-independent measurement of the matter-radiation equality scale in DESI 2024}",
      journal = {\prd},
         year = 2025,
        month = sep,
       volume = {112},
       number = {6},
          eid = {063553},
        pages = {063553},
          doi = {10.1103/yqm1-ybbv},
archivePrefix = {arXiv},
       eprint = {2505.16153},
 primaryClass = {astro-ph.CO},
       adsurl = {https://ui.adsabs.harvard.edu/abs/2025PhRvD.112f3553B}
}

@ARTICLE{2025OJAp....8E..42A,
       author = {{Alonso}, David and {Hetmantsev}, Oleksandr and {Fabbian}, Giulio and {Slosar}, Anze and {Storey-Fisher}, Kate},
        title = "{Measurement of the power spectrum turnover scale from the cross-correlation between CMB lensing and Quaia}",
      journal = {The Open Journal of Astrophysics},
         year = 2025,
        month = apr,
       volume = {8},
          eid = {42},
        pages = {42},
          doi = {10.33232/001c.136891},
archivePrefix = {arXiv},
       eprint = {2410.24134},
 primaryClass = {astro-ph.CO},
       adsurl = {https://ui.adsabs.harvard.edu/abs/2025OJAp....8E..42A}
}

@ARTICLE{2005MNRAS.363.1329B,
       author = {{Blake}, Chris and {Bridle}, Sarah},
        title = "{Cosmology with photometric redshift surveys}",
      journal = {\mnras},
         year = 2005,
        month = nov,
       volume = {363},
       number = {4},
        pages = {1329-1348},
          doi = {10.1111/j.1365-2966.2005.09526.x},
archivePrefix = {arXiv},
       eprint = {astro-ph/0411713},
 primaryClass = {astro-ph},
       adsurl = {https://ui.adsabs.harvard.edu/abs/2005MNRAS.363.1329B}
}

@ARTICLE{2013MNRAS.429.1902P,
       author = {{Poole}, Gregory B. and {Blake}, Chris and {Parkinson}, David and {Brough}, Sarah and {Colless}, Matthew and {Contreras}, Carlos and {Couch}, Warrick and {Croton}, Darren J. and {Croom}, Scott and {Davis}, Tamara and {Drinkwater}, Michael J. and {Forster}, Karl and {Gilbank}, David and {Gladders}, Mike and {Glazebrook}, Karl and {Jelliffe}, Ben and {Jurek}, Russell J. and {Li}, I.-hui and {Madore}, Barry and {Martin}, D. Christopher and {Pimbblet}, Kevin and {Pracy}, Michael and {Sharp}, Rob and {Wisnioski}, Emily and {Woods}, David and {Wyder}, Ted K. and {Yee}, H.~K.~C.},
        title = "{The WiggleZ Dark Energy Survey: probing the epoch of radiation domination using large-scale structure}",
      journal = {\mnras},
         year = 2013,
        month = mar,
       volume = {429},
       number = {3},
        pages = {1902-1912},
          doi = {10.1093/mnras/sts431},
archivePrefix = {arXiv},
       eprint = {1211.5605},
 primaryClass = {astro-ph.CO},
       adsurl = {https://ui.adsabs.harvard.edu/abs/2013MNRAS.429.1902P}
}

@ARTICLE{2023MNRAS.524.2463B,
       author = {{Bahr-Kalus}, Benedict and {Parkinson}, David and {Mueller}, Eva-Maria},
        title = "{Measurement of the matter-radiation equality scale using the extended baryon oscillation spectroscopic survey quasar sample}",
      journal = {\mnras},
         year = 2023,
        month = sep,
       volume = {524},
       number = {2},
        pages = {2463-2476},
          doi = {10.1093/mnras/stad1867},
archivePrefix = {arXiv},
       eprint = {2302.07484},
 primaryClass = {astro-ph.CO},
       adsurl = {https://ui.adsabs.harvard.edu/abs/2023MNRAS.524.2463B}
}

@ARTICLE{de_lucia_2007,
       author = {{De Lucia}, Gabriella and {Blaizot}, J{\'e}r{\'e}my},
        title = "{The hierarchical formation of the brightest cluster galaxies}",
      journal = {\mnras},
         year = 2007,
        month = feb,
       volume = {375},
       number = {1},
        pages = {2-14},
          doi = {10.1111/j.1365-2966.2006.11287.x},
archivePrefix = {arXiv},
       eprint = {astro-ph/0606519},
 primaryClass = {astro-ph},
       adsurl = {https://ui.adsabs.harvard.edu/abs/2007MNRAS.375....2D}
}

@ARTICLE{de_lucia_2014,
       author = {{De Lucia}, Gabriella and {Tornatore}, Luca and {Frenk}, Carlos S. and {Helmi}, Amina and {Navarro}, Julio F. and {White}, Simon D.~M.},
        title = "{Elemental abundances in Milky Way-like galaxies from a hierarchical galaxy formation model}",
      journal = {\mnras},
         year = 2014,
        month = nov,
       volume = {445},
       number = {1},
        pages = {970-987},
          doi = {10.1093/mnras/stu1752},
archivePrefix = {arXiv},
       eprint = {1407.7867},
 primaryClass = {astro-ph.GA},
       adsurl = {https://ui.adsabs.harvard.edu/abs/2014MNRAS.445..970D}
}

@ARTICLE{de_lucia_2024,
       author = {{De Lucia}, Gabriella and {Fontanot}, Fabio and {Xie}, Lizhi and {Hirschmann}, Michaela},
        title = "{Tracing the quenching journey across cosmic time}",
      journal = {\aap},
         year = 2024,
        month = jul,
       volume = {687},
          eid = {A68},
        pages = {A68},
          doi = {10.1051/0004-6361/202349045},
archivePrefix = {arXiv},
       eprint = {2401.06211},
 primaryClass = {astro-ph.GA},
       adsurl = {https://ui.adsabs.harvard.edu/abs/2024A&A...687A..68D}
}

@ARTICLE{xie_2017,
       author = {{Xie}, Lizhi and {De Lucia}, Gabriella and {Hirschmann}, Michaela and {Fontanot}, Fabio and {Zoldan}, Anna},
        title = "{H$_{2}$-based star formation laws in hierarchical models of galaxy formation}",
      journal = {\mnras},
         year = 2017,
        month = jul,
       volume = {469},
       number = {1},
        pages = {968-993},
          doi = {10.1093/mnras/stx889},
archivePrefix = {arXiv},
       eprint = {1611.09372},
 primaryClass = {astro-ph.GA},
       adsurl = {https://ui.adsabs.harvard.edu/abs/2017MNRAS.469..968X}
}

@ARTICLE{xie_2020,
       author = {{Xie}, Lizhi and {De Lucia}, Gabriella and {Hirschmann}, Michaela and {Fontanot}, Fabio},
        title = "{The influence of environment on satellite galaxies in the GAEA semi-analytic model}",
      journal = {\mnras},
         year = 2020,
        month = nov,
       volume = {498},
       number = {3},
        pages = {4327-4344},
          doi = {10.1093/mnras/staa2370},
archivePrefix = {arXiv},
       eprint = {2003.12757},
 primaryClass = {astro-ph.GA},
       adsurl = {https://ui.adsabs.harvard.edu/abs/2020MNRAS.498.4327X}
}

@ARTICLE{fontanot_2020,
       author = {{Fontanot}, Fabio and {De Lucia}, Gabriella and {Hirschmann}, Michaela and {Xie}, Lizhi and {Monaco}, Pierluigi and {Menci}, Nicola and {Fiore}, Fabrizio and {Feruglio}, Chiara and {Cristiani}, Stefano and {Shankar}, Francesco},
        title = "{The rise of active galactic nuclei in the galaxy evolution and assembly semi-analytic model}",
      journal = {\mnras},
         year = 2020,
        month = aug,
       volume = {496},
       number = {3},
        pages = {3943-3960},
          doi = {10.1093/mnras/staa1716},
archivePrefix = {arXiv},
       eprint = {2002.10576},
 primaryClass = {astro-ph.CO},
       adsurl = {https://ui.adsabs.harvard.edu/abs/2020MNRAS.496.3943F}
}

@ARTICLE{fontanot_2025,
       author = {{Fontanot}, Fabio and {De Lucia}, Gabriella and {Xie}, Lizhi and {Hirschmann}, Michaela and {Baugh}, Carlton and {Helly}, John C.},
        title = "{Galaxy assembly and evolution in the P-Millennium simulation: Galaxy clustering}",
      journal = {\aap},
         year = 2025,
        month = jul,
       volume = {699},
          eid = {A108},
        pages = {A108},
          doi = {10.1051/0004-6361/202452029},
archivePrefix = {arXiv},
       eprint = {2409.02194},
 primaryClass = {astro-ph.GA},
       adsurl = {https://ui.adsabs.harvard.edu/abs/2025A&A...699A.108F}
}

@ARTICLE{hirschmann_2016,
       author = {{Hirschmann}, Michaela and {De Lucia}, Gabriella and {Fontanot}, Fabio},
        title = "{Galaxy assembly, stellar feedback and metal enrichment: the view from the GAEA model}",
      journal = {\mnras},
         year = 2016,
        month = sep,
       volume = {461},
       number = {2},
        pages = {1760-1785},
          doi = {10.1093/mnras/stw1318},
archivePrefix = {arXiv},
       eprint = {1512.04531},
 primaryClass = {astro-ph.GA},
       adsurl = {https://ui.adsabs.harvard.edu/abs/2016MNRAS.461.1760H}
}

@misc{mayor_2026,
  author       = {{Mayor}, Jo\"{e}l and {Spinelli}, Marta and {De Lucia}, Gabriella and {Yates}, Robert and {Refregier}, Alexandre and {Fontanot}, Fabio and {Xie}, Lizhi and {Hirschmann}, Michaela},
  year         = {2026},
  title        = {Simulations of the 21cm emission line for upcoming large-scale HI galaxy surveys},
  eprint       = {2602.21058},
  archivePrefix= {arXiv},
  primaryClass = {astro-ph.CO},
  url          = {https://arxiv.org/abs/2602.21058}, 
}

@ARTICLE{2022A&A...662A.112E,
       author = {{Euclid Collaboration} and {Scaramella}, R. and {Amiaux}, J. and {Mellier}, Y. and {Burigana}, C. and {Carvalho}, C.~S. and {Cuillandre}, J.-C. and {Da Silva}, A. and {Derosa}, A. and {Dinis}, J. and {Maiorano}, E. and {Maris}, M. and {Tereno}, I. and {Laureijs}, R. and {Boenke}, T. and {Buenadicha}, G. and {Dupac}, X. and {Gaspar Venancio}, L.~M. and {G{\'o}mez-{\'A}lvarez}, P. and {Hoar}, J. and {Lorenzo Alvarez}, J. and {Racca}, G.~D. and {Saavedra-Criado}, G. and {Schwartz}, J. and {Vavrek}, R. and {Schirmer}, M. and {Aussel}, H. and {Azzollini}, R. and {Cardone}, V.~F. and {Cropper}, M. and {Ealet}, A. and {Garilli}, B. and {Gillard}, W. and {Granett}, B.~R. and {Guzzo}, L. and {Hoekstra}, H. and {Jahnke}, K. and {Kitching}, T. and {Maciaszek}, T. and {Meneghetti}, M. and {Miller}, L. and {Nakajima}, R. and {Niemi}, S.~M. and {Pasian}, F. and {Percival}, W.~J. and {Pottinger}, S. and {Sauvage}, M. and {Scodeggio}, M. and {Wachter}, S. and {Zacchei}, A. and {Aghanim}, N. and {Amara}, A. and {Auphan}, T. and {Auricchio}, N. and {Awan}, S. and {Balestra}, A. and {Bender}, R. and {Bodendorf}, C. and {Bonino}, D. and {Branchini}, E. and {Brau-Nogue}, S. and {Brescia}, M. and {Candini}, G.~P. and {Capobianco}, V. and {Carbone}, C. and {Carlberg}, R.~G. and {Carretero}, J. and {Casas}, R. and {Castander}, F.~J. and {Castellano}, M. and {Cavuoti}, S. and {Cimatti}, A. and {Cledassou}, R. and {Congedo}, G. and {Conselice}, C.~J. and {Conversi}, L. and {Copin}, Y. and {Corcione}, L. and {Costille}, A. and {Courbin}, F. and {Degaudenzi}, H. and {Douspis}, M. and {Dubath}, F. and {Duncan}, C.~A.~J. and {Dusini}, S. and {Farrens}, S. and {Ferriol}, S. and {Fosalba}, P. and {Fourmanoit}, N. and {Frailis}, M. and {Franceschi}, E. and {Franzetti}, P. and {Fumana}, M. and {Gillis}, B. and {Giocoli}, C. and {Grazian}, A. and {Grupp}, F. and {Haugan}, S.~V.~H. and {Holmes}, W. and {Hormuth}, F. and {Hudelot}, P. and {Kermiche}, S. and {Kiessling}, A. and {Kilbinger}, M. and {Kohley}, R. and {Kubik}, B. and {K{\"u}mmel}, M. and {Kunz}, M. and {Kurki-Suonio}, H. and {Lahav}, O. and {Ligori}, S. and {Lilje}, P.~B. and {Lloro}, I. and {Mansutti}, O. and {Marggraf}, O. and {Markovic}, K. and {Marulli}, F. and {Massey}, R. and {Maurogordato}, S. and {Melchior}, M. and {Merlin}, E. and {Meylan}, G. and {Mohr}, J.~J. and {Moresco}, M. and {Morin}, B. and {Moscardini}, L. and {Munari}, E. and {Nichol}, R.~C. and {Padilla}, C. and {Paltani}, S. and {Peacock}, J. and {Pedersen}, K. and {Pettorino}, V. and {Pires}, S. and {Poncet}, M. and {Popa}, L. and {Pozzetti}, L. and {Raison}, F. and {Rebolo}, R. and {Rhodes}, J. and {Rix}, H.-W. and {Roncarelli}, M. and {Rossetti}, E. and {Saglia}, R. and {Schneider}, P. and {Schrabback}, T. and {Secroun}, A. and {Seidel}, G. and {Serrano}, S. and {Sirignano}, C. and {Sirri}, G. and {Skottfelt}, J. and {Stanco}, L. and {Starck}, J.~L. and {Tallada-Cresp{\'\i}}, P. and {Tavagnacco}, D. and {Taylor}, A.~N. and {Teplitz}, H.~I. and {Toledo-Moreo}, R. and {Torradeflot}, F. and {Trifoglio}, M. and {Valentijn}, E.~A. and {Valenziano}, L. and {Verdoes Kleijn}, G.~A. and {Wang}, Y. and {Welikala}, N. and {Weller}, J. and {Wetzstein}, M. and {Zamorani}, G. and {Zoubian}, J. and {Andreon}, S. and {Baldi}, M. and {Bardelli}, S. and {Boucaud}, A. and {Camera}, S. and {Di Ferdinando}, D. and {Fabbian}, G. and {Farinelli}, R. and {Galeotta}, S. and {Graci{\'a}-Carpio}, J. and {Maino}, D. and {Medinaceli}, E. and {Mei}, S. and {Neissner}, C. and {Polenta}, G. and {Renzi}, A. and {Romelli}, E. and {Rosset}, C. and {Sureau}, F. and {Tenti}, M. and {Vassallo}, T. and {Zucca}, E. and {Baccigalupi}, C. and {Balaguera-Antol{\'\i}nez}, A. and {Battaglia}, P. and {Biviano}, A. and {Borgani}, S. and {Bozzo}, E. and {Cabanac}, R. and {Cappi}, A.},
        title = "{Euclid preparation. I. The Euclid Wide Survey}",
      journal = {\aap},
         year = 2022,
        month = jun,
       volume = {662},
          eid = {A112},
        pages = {A112},
          doi = {10.1051/0004-6361/202141938},
archivePrefix = {arXiv},
       eprint = {2108.01201},
 primaryClass = {astro-ph.CO},
       adsurl = {https://ui.adsabs.harvard.edu/abs/2022A&A...662A.112E}
}

@ARTICLE{2020MNRAS.498.2354R,
       author = {{Ross}, Ashley J. and {Bautista}, Julian and {Tojeiro}, Rita and {Alam}, Shadab and {Bailey}, Stephen and {Burtin}, Etienne and {Comparat}, Johan and {Dawson}, Kyle S. and {de Mattia}, Arnaud and {du Mas des Bourboux}, H{\'e}lion and {Gil-Mar{\'\i}n}, H{\'e}ctor and {Hou}, Jiamin and {Kong}, Hui and {Lyke}, Brad W. and {Mohammad}, Faizan G. and {Moustakas}, John and {Mueller}, Eva-Maria and {Myers}, Adam D. and {Percival}, Will J. and {Raichoor}, Anand and {Rezaie}, Mehdi and {Seo}, Hee-Jong and {Smith}, Alex and {Tinker}, Jeremy L. and {Zarrouk}, Pauline and {Zhao}, Cheng and {Zhao}, Gong-Bo and {Bizyaev}, Dmitry and {Brinkmann}, Jonathan and {Brownstein}, Joel R. and {Rosell}, Aurelio Carnero and {Chabanier}, Sol{\`e}ne and {Choi}, Peter D. and {Chuang}, Chia-Hsun and {Cruz-Gonzalez}, Irene and {de la Macorra}, Axel and {de la Torre}, Sylvain and {Escoffier}, Stephanie and {Fromenteau}, Sebastien and {Higley}, Alexandra and {Jullo}, Eric and {Kneib}, Jean-Paul and {McLane}, Jacob N. and {Mu{\~n}oz-Guti{\'e}rrez}, Andrea and {Neveux}, Richard and {Newman}, Jeffrey A. and {Nitschelm}, Christian and {Palanque-Delabrouille}, Nathalie and {Paviot}, Romain and {Pullen}, Anthony R. and {Rossi}, Graziano and {Ruhlmann-Kleider}, Vanina and {Schneider}, Donald P. and {Maga{\~n}a}, Mariana Vargas and {Vivek}, M. and {Zhang}, Yucheng},
        title = "{The Completed SDSS-IV extended Baryon Oscillation Spectroscopic Survey: Large-scale structure catalogues for cosmological analysis}",
      journal = {\mnras},
         year = 2020,
        month = oct,
       volume = {498},
       number = {2},
        pages = {2354-2371},
          doi = {10.1093/mnras/staa2416},
archivePrefix = {arXiv},
       eprint = {2007.09000},
 primaryClass = {astro-ph.CO},
       adsurl = {https://ui.adsabs.harvard.edu/abs/2020MNRAS.498.2354R}
}

@ARTICLE{2025arXiv250314745D,
       author = {{DESI Collaboration} and {Abdul-Karim}, M. and {Adame}, A.~G. and {Aguado}, D. and {Aguilar}, J. and {Ahlen}, S. and {Alam}, S. and {Aldering}, G. and {Alexander}, D.~M. and {Alfarsy}, R. and {Allen}, L. and {Allende Prieto}, C. and {Alves}, O. and {Anand}, A. and {Andrade}, U. and {Armengaud}, E. and {Avila}, S. and {Aviles}, A. and {Awan}, H. and {Bailey}, S. and {Baleato Lizancos}, A. and {Ballester}, O. and {Bault}, A. and {Bautista}, J. and {BenZvi}, S. and {Beraldo e Silva}, L. and {Bermejo-Climent}, J.~R. and {Beutler}, F. and {Bianchi}, D. and {Blake}, C. and {Blum}, R. and {Bolton}, A.~S. and {Bonici}, M. and {Brieden}, S. and {Brodzeller}, A. and {Brooks}, D. and {Buckley-Geer}, E. and {Burtin}, E. and {Canning}, R. and {Carnero Rosell}, A. and {Carr}, A. and {Carrilho}, P. and {Casas}, L. and {Castander}, F.~J. and {Cereskaite}, R. and {Cervantes-Cota}, J.~L. and {Chaussidon}, E. and {Chaves-Montero}, J. and {Chen}, S. and {Chen}, X. and {Claybaugh}, T. and {Cole}, S. and {Cooper}, A.~P. and {Cousinou}, M.-C. and {Cuceu}, A. and {Davis}, T.~M. and {Dawson}, K.~S. and {de Belsunce}, R. and {de la Cruz}, R. and {de la Macorra}, A. and {de Mattia}, A. and {Deiosso}, N. and {Della Costa}, J. and {Demina}, R. and {Demirbozan}, U. and {DeRose}, J. and {Dey}, A. and {Dey}, B. and {Ding}, J. and {Ding}, Z. and {Doel}, P. and {Douglass}, K. and {Dowicz}, M. and {Ebina}, H. and {Edelstein}, J. and {Eisenstein}, D.~J. and {Elbers}, W. and {Emas}, N. and {Escoffier}, S. and {Fagrelius}, P. and {Fan}, X. and {Fanning}, K. and {Fawcett}, V.~A. and {Fern'andez-Garc'ia}, E. and {Ferraro}, S. and {Findlay}, N. and {Font-Ribera}, A. and {Forero-Romero}, J.~E. and {Forero-S'anchez}, D. and {Frenk}, C.~S. and {G''ansicke}, B.~T. and {Galbany}, L. and {Garc'ia-Bellido}, J. and {Garcia-Quintero}, C. and {Garrison}, L.~H. and {Gaztanaga}, E. and {Gil-Mar'in}, H. and {Gnedin}, O.~Y. and {Gontcho}, S. Gontcho A and {Gonzalez-Morales}, A.~X. and {Gonzalez-Perez}, V. and {Gordon}, C. and {Graur}, O. and {Green}, D. and {Gruen}, D. and {Gsponer}, R. and {Guandalin}, C. and {Gutierrez}, G. and {Guy}, J. and {Hahn}, C. and {Han}, J.~J. and {Han}, J. and {He}, S. and {Herrera-Alcantar}, H.~K. and {Honscheid}, K. and {Hou}, J. and {Howlett}, C. and {Huterer}, D. and {Irv\{s\}iv\{c\}}, V. and {Ishak}, M. and {Jacques}, A. and {Jimenez}, J. and {Jing}, Y.~P. and {Joachimi}, B. and {Joudaki}, S. and {Joyce}, R. and {Jullo}, E. and {Juneau}, S. and {Karac\{c\}ayl\{i\}}, N.~G. and {Karim}, T. and {Kehoe}, R. and {Kent}, S. and {Khederlarian}, A. and {Kirkby}, D. and {Kisner}, T. and {Kitaura}, F.-S. and {Kizhuprakkat}, N. and {Kong}, H. and {Koposov}, S.~E. and {Kremin}, A. and {Krolewski}, A. and {Lahav}, O. and {Lai}, Y. and {Lamman}, C. and {Lan}, T.-W. and {Landriau}, M. and {Lang}, D. and {Lange}, J.~U. and {Lasker}, J. and {Le Goff}, J.~M. and {Le Guillou}, L. and {Leauthaud}, A. and {Levi}, M.~E. and {Li}, S. and {Li}, T.~S. and {Lodha}, K. and {Lokken}, M. and {Luo}, Y. and {Magneville}, C. and {Manera}, M. and {Manser}, C.~J. and {Margala}, D. and {Martini}, P. and {Maus}, M. and {McCullough}, J. and {McDonald}, P. and {Medina}, G.~E. and {Medina-Varela}, L. and {Meisner}, A. and {Mena-Fern'andez}, J. and {Menegas}, A. and {Mezcua}, M. and {Miquel}, R. and {Montero-Camacho}, P. and {Moon}, J. and {Moustakas}, J. and {Munoz-Guti'errez}, A. and {Munoz-Santos}, D. and {Myers}, A.~D. and {Myles}, J. and {Nadathur}, S. and {Najita}, J. and {Napolitano}, L. and {Newman}, J.~A. and {Nikakhtar}, F. and {Nikutta}, R. and {Niz}, G. and {Noriega}, H.~E. and {Padmanabhan}, N. and {Paillas}, E. and {Palanque-Delabrouille}, N. and {Palmese}, A. and {Pan}, J. and {Pan}, Z. and {Parkinson}, D. and {Peacock}, J. and {Percival}, W.~J. and {P'erez-Fern'andez}, A. and {P'erez-R`afols}, I. and {Peterson}, P.},
        title = "{Data Release 1 of the Dark Energy Spectroscopic Instrument}",
      journal = {arXiv e-prints},
         year = 2025,
        month = mar,
          eid = {arXiv:2503.14745},
        pages = {arXiv:2503.14745},
          doi = {10.48550/arXiv.2503.14745},
archivePrefix = {arXiv},
       eprint = {2503.14745},
 primaryClass = {astro-ph.CO},
       adsurl = {https://ui.adsabs.harvard.edu/abs/2025arXiv250314745D}
}

@ARTICLE{2006ApJ...647..201C,
       author = {{Conroy}, Charlie and {Wechsler}, Risa H. and {Kravtsov}, Andrey V.},
        title = "{Modeling Luminosity-dependent Galaxy Clustering through Cosmic Time}",
      journal = {\apj},
         year = 2006,
        month = aug,
       volume = {647},
       number = {1},
        pages = {201-214},
          doi = {10.1086/503602},
archivePrefix = {arXiv},
       eprint = {astro-ph/0512234},
 primaryClass = {astro-ph},
       adsurl = {https://ui.adsabs.harvard.edu/abs/2006ApJ...647..201C}
}

@INCOLLECTION{2010dken.book..246B,
       author = {{Bassett}, Bruce and {Hlozek}, Renee},
        title = "{Baryon acoustic oscillations}",
    booktitle = {Dark Energy},
         year = 2010,
       editor = {{Ruiz-Lapuente}, Pilar},
        pages = {246},
          doi = {10.48550/arXiv.0910.5224},
       adsurl = {https://ui.adsabs.harvard.edu/abs/2010dken.book..246B}
}

@ARTICLE{2003ApJ...598..720S,
       author = {{Seo}, Hee-Jong and {Eisenstein}, Daniel J.},
        title = "{Probing Dark Energy with Baryonic Acoustic Oscillations from Future Large Galaxy Redshift Surveys}",
      journal = {\apj},
         year = 2003,
        month = dec,
       volume = {598},
       number = {2},
        pages = {720-740},
          doi = {10.1086/379122},
archivePrefix = {arXiv},
       eprint = {astro-ph/0307460},
 primaryClass = {astro-ph},
       adsurl = {https://ui.adsabs.harvard.edu/abs/2003ApJ...598..720S}
}

@ARTICLE{2007ApJ...664..660E,
       author = {{Eisenstein}, Daniel J. and {Seo}, Hee-Jong and {White}, Martin},
        title = "{On the Robustness of the Acoustic Scale in the Low-Redshift Clustering of Matter}",
      journal = {\apj},
         year = 2007,
        month = aug,
       volume = {664},
       number = {2},
        pages = {660-674},
          doi = {10.1086/518755},
archivePrefix = {arXiv},
       eprint = {astro-ph/0604361},
 primaryClass = {astro-ph},
       adsurl = {https://ui.adsabs.harvard.edu/abs/2007ApJ...664..660E}
}

@ARTICLE{1987MNRAS.227....1K,
       author = {{Kaiser}, Nick},
        title = "{Clustering in real space and in redshift space}",
      journal = {\mnras},
         year = 1987,
        month = jul,
       volume = {227},
        pages = {1-21},
          doi = {10.1093/mnras/227.1.1},
       adsurl = {https://ui.adsabs.harvard.edu/abs/1987MNRAS.227....1K}
}

@ARTICLE{2002ApJ...575..587B,
       author = {{Berlind}, Andreas A. and {Weinberg}, David H.},
        title = "{The Halo Occupation Distribution: Toward an Empirical Determination of the Relation between Galaxies and Mass}",
      journal = {\apj},
         year = 2002,
        month = aug,
       volume = {575},
       number = {2},
        pages = {587-616},
          doi = {10.1086/341469},
archivePrefix = {arXiv},
       eprint = {astro-ph/0109001},
 primaryClass = {astro-ph},
       adsurl = {https://ui.adsabs.harvard.edu/abs/2002ApJ...575..587B}
}

@ARTICLE{2018PhR...733....1D,
       author = {{Desjacques}, Vincent and {Jeong}, Donghui and {Schmidt}, Fabian},
        title = "{Large-scale galaxy bias}",
      journal = {\physrep},
         year = 2018,
        month = feb,
       volume = {733},
        pages = {1-193},
          doi = {10.1016/j.physrep.2017.12.002},
archivePrefix = {arXiv},
       eprint = {1611.09787},
 primaryClass = {astro-ph.CO},
       adsurl = {https://ui.adsabs.harvard.edu/abs/2018PhR...733....1D}
}

@ARTICLE{2023OJAp....6E..39A,
       author = {{Asgari}, Marika and {Mead}, Alexander J. and {Heymans}, Catherine},
        title = "{The halo model for cosmology: a pedagogical review}",
      journal = {The Open Journal of Astrophysics},
         year = 2023,
        month = nov,
       volume = {6},
          eid = {39},
        pages = {39},
          doi = {10.21105/astro.2303.08752},
archivePrefix = {arXiv},
       eprint = {2303.08752},
 primaryClass = {astro-ph.CO},
       adsurl = {https://ui.adsabs.harvard.edu/abs/2023OJAp....6E..39A}
}

@INPROCEEDINGS{1991ASPC...19..428W,
       author = {{Wilkinson}, P.~N.},
        title = "{The Hydrogen Array}",
    booktitle = {IAU Colloquium 131: Radio Interferometry. Theory, Techniques, and Applications},
         year = 1991,
       editor = {{Cornwell}, T.~J. and {Perley}, R.~A.},
       series = {Astronomical Society of the Pacific Conference Series},
       volume = {19},
        month = jan,
        pages = {428-432},
       adsurl = {https://ui.adsabs.harvard.edu/abs/1991ASPC...19..428W}
}

@ARTICLE{2025JCAP...10..071L,
       author = {{Lai}, Y. and {Howlett}, C. and {Davis}, T.~M.},
        title = "{Can a multi-tracer approach improve the constraints on the turnover scale at low redshift?}",
      journal = {\jcap},
         year = 2025,
        month = oct,
       volume = {2025},
       number = {10},
          eid = {071},
        pages = {071},
          doi = {10.1088/1475-7516/2025/10/071},
archivePrefix = {arXiv},
       eprint = {2507.11823},
 primaryClass = {astro-ph.CO},
       adsurl = {https://ui.adsabs.harvard.edu/abs/2025JCAP...10..071L}
}

\end{document}